# Jumps: Enhancing hop-count positioning in sensor networks using multiple coordinates

Farid Benbadis [a] Jean-Jacques Puig [b] Marcelo Dias de Amorim [a]
Claude Chaudet [b] Timur Friedman [a] David Simplot-Ryl [c]

[a] *Université Pierre et Marie Curie – Paris VI, LIP6-UMR 7606 CNRS, Paris, France*

[b] *GET/ENST; LTCI-UMR 5141 CNRS, Paris, France*

[c] *IRCICA/LIFL, Univ. Lille 1-UMR 8022 CNRS, INRIA Futurs, Lille, France*

**Abstract**

Positioning systems in self-organizing networks generally rely on measurements such as delay and received signal strength, which may be difficult to obtain and often require dedicated equipment. An alternative to such approaches is to use simple connectivity information, that is, the presence or absence of a link between any pair of nodes, and to extend it to hop-counts, in order to obtain an approximate coordinate system. Such an approximation is sufficient for a large number of applications, such as routing. In this paper, we propose Jumps, a positioning system for those self-organizing networks in which other types of (exact) positioning systems cannot be used or are deemed to be too costly. Jumps builds a multiple coordinate system based solely on nodes' neighborhood knowledge. Jumps is interesting in the context of wireless sensor networks, as it neither requires additional embedded equipment nor relies on any node's capabilities. While other approaches use only three hop-count measurements to infer the position of a node, Jumps uses an arbitrary number. We observe that an increase in the number of measurements leads to an improvement in the localization process, without requiring a high dense environment. We show through simulations that Jumps, when compared with existing approaches, reduces the number of nodes sharing the same coordinates, which paves the way for functions such as position-based routing.

*Email addresses:* `Farid.Benbadis@lip6.fr` (Farid Benbadis), `JeanJacques.Puig@enst.fr` (Jean-Jacques Puig), `Marcelo.Amorim@lip6.fr` (Marcelo Dias de Amorim), `Claude.Chaudet@enst.fr` (Claude Chaudet),

## 1 Introduction

Wireless sensor networks are composed of cooperative elements that acquire measurements of physical phenomena and send the resulting data to a collector node, called sink. Sensors are usually characterized by their small size, light weight, low power consumption, and limited memory and energy. In such a context, any functionality introduced into the network must be carefully designed and deployed.

While some scenarios allow for manual deployment of sensors (e.g., structure monitoring), in others, sensors deployment might be disordered or even random (e.g., a battlefield). In this second type of scenario, the network needs to self-organize, in order to provide nodes with identifiers and to build routing paths. Knowing the geographic positions of the nodes can be of great utility in self-organizing networks. For instance, node positions can be used to facilitate and increase the efficiency of routing protocols [1,2]. In sensor networks, positioning is particularly important, since gathering physical measurements without attaching coordinates to the collected data might be of little use. In this paper, we focus on the process of attributing coordinates to nodes in sensor networks.

The state of the art in absolute positioning systems is dominated by satellite-based methods, like GPS [3], Glonass [4], and the upcoming Galileo [5]. Yet, equipping each sensor node with a satellite receiver might neither be feasible nor always useful. Such a receiver would increase the cost of the sensor network, the size and weight of the individual sensors, as well as the overall power consumption, which is in contradiction with the objectives of most sensor network scenarios.

Some solutions propose to equip a subset of the nodes with a satellite receiver and use these nodes as references, also called *landmarks*, to infer the positions of other nodes by using signal strength measurements and deducing an approximation of their Euclidean distances to each of these landmarks using radio propagation laws. While this situation can lead to accurate positioning of the nodes in an outdoor scenario, it is much less efficient (or even unfeasible) in indoor or underground environments.

To compute relative positions of nodes, for example when satellite receivers are not available, or simply to enhance satellite measurements, nodes can use different types of information. *Received signal strength* (RSS) can be converted into distances when the propagation law is uniform and known, either directly or using multiple signals and *time difference of arrival* (TDoA) [6].

`Timur.Friedman@upmc.fr` (Timur Friedman), `David.Simplot@lifl.fr` (David Simplot-Ryl).

Positions can then be inferred using trilateration techniques. Directional antennae can also be used to triangulate relative positions [7]. In spite of their accuracy, these techniques might be unusable because of the dedicated and costly equipment they require. Using only connectivity information is algorithmically simpler and, despite its relative simplicity, can still lead to usable positions. Rao et al. propose [8] to discover border nodes, compute their coordinates, and finally infer, through a relaxation method, their coordinates to other nodes. Benbadis et al., authors of the present paper, propose in a prior work [9], as Caruso et al. [10], to use hop distances to landmarks to compute nodes coordinates. Both works have shown that using hop-counts is a sufficient metric to implement positioning systems in a number of applications.

The use of landmarks is not exclusive to the context of self-organizing networks. Geoping [11] and GeoLIM [12] use well-known landmarks to attribute coordinates to Internet hosts. In the latter case, GeoLIM aims at providing the geographic location of an Internet host using its IP address. GeoLIM transforms delay measurements between landmarks at well known locations and a target host into geographic distance constraints. Like GPS, it uses multilateration to estimate the geographic location of the target host. While three landmarks are sufficient to estimate a host position, GeoLIM uses up to 74 landmarks to increase the accuracy of its location estimations. We can learn from the Internet experience and transpose the use of multiple landmarks to the context of sensor networks. However, less information is available in our context, as it lacks preliminary location information. Moreover, delays are not accurate in the wireless context due to the unpredictable nature of the medium.

The idea of using multiple landmarks and hop-counts between nodes and landmarks seems to be suitable for positioning in sensor networks too. Yet prior work [9,10] has been constrained to three landmarks. The question we address in this paper is: what improvement can be obtained by the use of a greater number of landmarks? To the best of our knowledge, no other study uses hop-counts with more than three landmarks, and so the potential benefits are unknown.

The idea behind Jumps, the system we propose here, is to construct a virtual coordinate system where each node uses its hop-count distances to each landmark as its coordinates. In the resulting system, each node has as many coordinates as there are landmarks. In order to evaluate the impact of this additional information on the accuracy of the system, we have performed a set of simulations. Prior work [9,10] concluded that increasing the node density, either by introducing more nodes or by increasing transmission range, improved the localization accuracy. Our simulation results show that increasing the number of landmarks leads to a similar improvement without an increase in the number of nodes. We also show, through a theoretical study, that using more landmarks instead of increasing the node density does not increase the energy consumption.

This paper is organized as follows. Section 2 presents prior work in positioning algorithms for sensor networks. Section 3 describes our multi-coordinate hop-count positioning algorithm. Section 4 describes the simulations we performed and analyzes the results. Section 5 analyzes the energy consumption of Jumps and compares it with other approaches' one. Section 6 draws conclusions and points to future work.

## 2 Positioning systems

Many positioning systems for sensor networks have been proposed in the literature. In this section, we briefly present the most important ones and focus more specifically on the solutions that use hop-counts as distances, as they are the most closely related to our proposal. All the solutions are measurement-based, and we classify algorithms into two distinct groups: signal-based approaches and connectivity-based approaches. While in signal-based approaches measurements are performed on the physical characteristics of each link or the angle of the received signal, connectivity-based approaches rely solely on the existence, or lack thereof, of a link between two nodes. In the following, we explain each of these groups.

### *2.1 Signal-based approaches*

This family of mechanisms uses the physical characteristics of links between pairs of nodes in order to estimate distance. The first class of methods uses the *received signal strength* (RSS). RSS is based on the principal that a radio signal between a sender and a receiver attenuates with an increase in distance. Bahl et al. [13] suggest the use of average received signal strength to estimate distances. To take into account the fact that, in the presence of obstacles, signals do not follow the free space attenuation model, they apply a wall attenuation factor [14]. The estimated distances are then used to locate a node by trilateration, i.e. by estimating the distances to three points of known coordinates and solving the equation corresponding to the intersection of the three corresponding circles to deduce the related node's position.

The *Ad hoc Positioning System* (APS) proposed by Niculescu and Nath [7] uses the capability of certain nodes to measure the *angle of arrival* (AoA) of incoming signals. They assume that some of the nodes know their exact positions via GPS, and that each node that joins the network is equipped with a directional antenna. Sharing location and angle information through a distance-vector-like information propagation mechanism allows non-GPS nodes to estimate their locations through trilateration.

Gustafsson and Gunnarsson use the *Time Difference Of Arrival* (TDoA) between multiple signals to compute a relative coordinate system [6]. In this approach, when a node sends a signal to two receivers, the correlation analysis of the received signals gives one hyperbolic function. With a larger number of receivers, more hyperbolic functions are obtained. Computing the intersection of these hyperbolas gives the estimated coordinates of the sending node.

Kwon et al. [15] propose a variation of TDoA to localize nodes in a sensor network using sound signals. This method assumes a topology with GPS and non-GPS nodes. GPS nodes, called anchor nodes, know their coordinates and are equipped with a compact loud speaker, called *sounder*. Non-GPS nodes have a microphone to receive sound signals from anchor nodes. Four anchor nodes, at least, should be used in order to create a 3-dimensional coordinate system. The TDoA service implemented here is used to measure distances needed for trilateration. The method uses the difference in arrival time of radio and sound signals from an anchor node to compute the distance.

Even if in the original TDoA method no specific equipment is required, a time synchronization is necessary in order to lead to good position estimation. Such synchronization requires specific devices, e.g. GPS, which makes this method difficult to implement in many situations, e.g. underground scenarios where GPS is not available.

In GPS-free [16], each node discovers its neighborhood and measures its distance as its delay to all of its one-hop neighbors. The list of all of a nodes' neighbors and their distances is sent to all of the nodes' neighbors. Every node then knows the identity of its one and two-hops neighbors and some of the distances between them. Considering itself as the center of its own coordinate system, a node is able to compute coordinates for every node among its two-hops neighbors. The global network coordinate system is then derived by translating each local coordinate systems into a global coordinate system.

For comprehensive surveys on these signal-based methods, please refer to Nordlund et al. [17] and Hightower et al. [18].

### *2.2 Connectivity-based approaches*

This section surveys positioning algorithms that do not make any assumption regarding node capabilities other than their ability to discover their direct neighbors.

Rao et al. [8] use only hop-count distance to achieve positioning in an ad hoc network. Their algorithm involves three steps. First, nodes evaluate their hop-count distance to a bootstrap node. The nodes that are at a maximum distance from the bootstrap node in their neighborhood are assumed to be on the boundary of the network. Second, each border node sends a beacon that contains a counter. This operation allows border nodes to discover their distance to all other border nodes and build a vector with these distances. Each border node then sends its distance vector to all the other border nodes. The resulting distance matrix is then used to compute, through a minimizing function, virtual coordinates for the border nodes. Finally, each node computes iteratively its coordinates as the average of its neighbors coordinates.

Although this method gives accurate coordinates, it leads to high overhead since too many network-wide floods are required. Such a method is not advisable for a sensor network, where energy efficiency is paramount and in which radio communications are an important source of energy consumption. The use of hop-count distance is, however, promising measurements to exploit since it does not require any specific material and is easy to implement.

GPS-Free-Free [9], proposed by the authors of the present paper, and Caruso et al.'s VCap [10] use this simple measurement to construct a coordinate system. In both methods, authors use three nodes as landmarks. These three nodes are similar to the other nodes and have no any additional capabilities. The hop-count distances to these landmarks form the basis for a coordinate system. While GPS-Free-Free constructs a coordinate system based on the trilateration of the three hop-count distances, VCap uses the distances directly as coordinates. We believe that VCap, since it uses the available information more fully, should give more accurate results with less computation than our own earlier proposal.

### *2.3 Injecting redundancy into the coordinates system with* Jumps

GPS-Free-Free and VCap both suffer from a problem: using hop-counts as metrics introduces discretization into the distance measurements. For instance, a node that is at a distance of three hops from a particular landmark is, under certain density assumptions, between two and three times the communication range of a node. Jumps, our proposal, allows further smoothing by increasing the number of landmarks. It relies on the same principles as GPS-Free-Free and VCap, i.e., it is based purely on the use of landmarks and hop-count distances. We show in this paper that the number of landmarks is a fundamental parameter for the accuracy of the positioning system.

## 3 Virtual positioning using $N$ coordinates

The algorithm presented here creates a virtual coordinate system for sensor networks. We consider a static, random network with no isolated nodes. [1] For each node, we call *neighbors* the set of nodes within its transmission range; we further assume links to be symmetric (if $a$ is the neighbor of $b$, then $b$ is the neighbor of $a$). A critical parameter for our evaluation, presented in Section 4, is the neighbor density, given by the average number of neighbors per node and further described in Section 4.2.1.

Our algorithm uses $N$ landmarks to attribute as many coordinates to each node. These landmarks may be either specific, dedicated nodes, or a set of nodes dynamically selected within the network. In this paper, we do not specifically address the problem of selecting which nodes should play the role of landmarks, and focus on the study of the efficiency of using more than three landmarks to build a virtual coordinate system. The landmark positions are, in this study, set on the perimeter of the network, regularly spaced. We opted for such a configuration for generality reasons. For instance, a number of landmark placement techniques rely on network-wide floods, and a specific flooding technique may have an impact on the results. By defining a regular structure inscribed in a circle to avoid any bias from specific landmark determination algorithm in our results.

Let $\mathcal{L} = \{L_1, L_2, \ldots, L_N\}$ be the set of $N$ landmarks chosen to build the coordinate system. For any node $n_i$, $\{x_{i_1}, x_{i_2}, \ldots, x_{i_N}\}$ denotes the coordinates of $n_i$ using JUMPS, where $x_{i_j}$ is the minimum hop-count distance, i.e., the lowest count of successive links, from $L_j$ to $n_i$. Thus, the underlying basis of our coordinate system is that each node uses as coordinates its hop-count distances to each of the $N$ landmarks. A positioning mechanism should enable nodes to learn these distances from the network.

JUMPS' positioning phase begins upon completion of network deployment. There are two steps. While we do not focus on specific methods for these steps, we propose a generic method for any situation. First, a node wakes up, and second, it obtains its coordinates.

We assume that a node, which may be the sink, is chosen to initiate the first step; any node can be assigned (*e.g.* through specific software configuration) with this task, regardless of its physical location. We denote such a node as $s$, which can be chosen periodically through a local random process.

[1] Precisely, for any two nodes $n_i$, $n_j$, there is always a succession of one or more links between $n_i$ and $n_j$.

Note that nodes do not need to be synchronized. Any node can promote itself as initiator, and multiple nodes may get promoted. The only condition to be respected is that the random promotion process be stopped once a `WAKE` message is received.

Once chosen, node $s$ generates and floods a wake up message, noted `WAKE`, which purpose is to trigger the hop-count distances discovery between landmarks and the remainder of the nodes. Upon reception of the `WAKE` message, landmark $L_1$ initiates the second step, which consists in making nodes learn their hop-count distances to $L_1$. To this end, $L_1$ generates a *Distance Discovery Message*, $\mathtt{DDM}_{L_1}$, which carries a counter initially set to 0, and floods it throughout the network. Nodes receiving this message increment the counter and forward the message, following a classical flooding algorithm. Clearly, a node does not forward a DDM if it has already received a copy of the same message with a lower or equal counter value. When the flood is over, each node knows its hop-count distance to $L_1$. The main goal of $\mathtt{DDM}_{L_1}$ is achieved.

At the same time, a second functionality of $\mathtt{DDM}_{L_1}$ has been obtained: it wakes up $L_2$, which, in turn, generates $\mathtt{DDM}_{L_2}$ in order to initiate the establishment of the second axis of the virtual coordinate system. The algorithm can be generalized in the following way. Any landmark $L_i$ is awaken by the reception of $\mathtt{DDM}_{L_{i-1}}$ and generates its own $\mathtt{DDM}_{L_i}$.

When the last landmark, $L_N$, has flooded $\mathtt{DDM}_{L_N}$, every node in the network knows its hop-count distance to every landmark. At this point, all nodes have their hop-count distance to landmarks as coordinates. If $d_1, d_2, \ldots, d_N$ denote the hop-distance from node $n$ to, respectively, landmarks $L_1$, $L_2$, ..., $L_N$, then $\{d_1, d_2, \ldots, d_N\}$ are the virtual coordinates of $n$ in this $N$-dimensional space.

Fig. 1 shows an example of the coordinate system attribution algorithm. In this figure, eight landmarks are used, and the virtual coordinates of $n$ are $\{3, 6, 7, 7, 8, 7, 6, 3\}$, according to its hop-distance to landmarks.

## 4 Numerical Analysis

In order to evaluate the efficiency of our solution, we performed a set of simulations. We simulated static sensor networks within which up to ten landmarks were deployed on the field's boundary. It is important to note that our simulator has been designed to provide further understanding on the behavior of the coordinate system, and does not intend to model realistic physical and MAC layers. The simulation environment, parameters, and measurements are described in the following.

### *4.1 Environment*

The simulated field is a disc area of 1000 meter radius, $R$. The node transmission range, $r$, is set to 50 meters.

With the exception of landmarks, node locations in the field are realizations of a uniformly distributed pseudo-random variable.

### *4.2 Parameters*

Prior work [10,9] focused on densely populated networks. In such cases, the correlation between hop-count and Euclidean distance increases with the density. This clearly helps hop-count based solutions, which explains why previous works limited their analysis to dense networks. In this paper, we are also interested in investigating sparse networks.

We adjust the density $D$ of nodes by choosing the number of nodes scattered in the field, $\mathcal{M}$.

Precisely:

$$D = \frac{\mathcal{M}}{\pi \times R^2} \cdot \qquad (1)$$

#### *4.2.1 Neighborhood density*

Nodes are uniformly distributed in the simulated area. Thus, the average number of nodes per coverage range $d_{cov}$ can be calculated as:

$$d_{cov} = D \times \pi \times r^2 = \frac{\mathcal{M}}{\pi \times R^2} \times \pi \times r^2 = \mathcal{M} \times \left(\frac{r}{R}\right)^2 \cdot \qquad (2)$$

However, in our simulations, for better comparison with Caruso et al.'s work [10], instead of using density, we consider the average number of neighbors per node, $d_{neig}$. We present results for $d_{neig}$ varying in the range from 10 to 50, in steps of 10. Basically, $d_{neig} = d_{cov} + 1$.

For $r = 50$ and $R = 1000$, we thus have the following relation between nodes population $\mathcal{M}$ and neighbor density $d_{neig}$:

$$\mathcal{M} = 400 \times (d_{neig} + 1) \cdot \qquad (3)$$

Consequently, our simulations use 4400, 8400, 12400, 16400 and 20400 nodes in order to simulate networks of respective neighbor densities of 10, 20, 30, 40 and 50.

### *4.2.2 Number of landmarks*

The number of landmarks is a crucial parameter in our work. We simulated topologies with a number of landmarks, $N$, varying from 3 to 10. Landmarks are placed on the border of the field, at equal angles from the center of the field (cf. Fig. 1).

## *4.3 Metrics*

In general, evaluation of positioning mechanisms is based on the error between the computed coordinates and the real ones. This is meaningless in our work because the coordinate system we build is purely virtual and uses $N$ coordinates rather than the 2 or 3 coordinates of physical space. Again, our goal is not to get to such an approximation, but to provide a virtual system that would help networking functionalities such as routing.

Instead, we focus on the following metrics (illustrated in Fig.2):

- **Zone**: following the definition by Caruso et al. [10], a *zone* is a set of two or more nodes sharing the same set of virtual coordinates. The use of hop-counts as a distance measure tends to favor coordinate collisions.
- **Intra-zone distance**: this parameter represents the average Euclidean distance, in the physical world, between two nodes belonging to the same zone.
- **Zone size**: the maximum distance between two nodes within the same zone.
- **Maximum zone size**: defines the size of the largest zone in the topology. It is equivalent to the maximum distance between any two nodes sharing the same virtual coordinates in the whole network. In a nutshell, this measure provides the maximum localization error in the network.
- **Number of nodes per zone**: the average population in a zone.
- **Number of zones**: the number of zones in the network.

For each scenario, defined by an instance of couple $(N, d_{neig})$, we simulated 1000 different topologies, and considered only completely connected networks. Curves in the figures are accompanied by 99,9% confidence intervals. In figures, distances such as *intra-zone distance*, *zone size* and *maximum zone size* were divided by $r$ and noted as factors of the radio range (reference value). Since histograms led to poor legibility, we preferred to represent data as a function of the number of landmarks.

### *4.4 Results*

In this section, we present the results of the simulations we performed. We have performed experimentations with forty scenarios, but for the sake of legibility, we show only nine representative curves in distribution figures. The other thirty-one scenarios can be obtained online [19].

#### *4.4.1 Zone size*

Fig. 3 presents the repartition diagrams of zone sizes for different numbers of landmarks and different densities. By separating the columns, we can compare the effect of density with a fixed number of landmarks. Observe that the distribution does not vary too much, whatever the number of landmarks. On the other hand, by separating the rows, we can compare the effect of the number of landmarks. We clearly see that increasing the number of landmarks reduces the number of zones with sizes larger than one hop.

We recall that zone sizes are represented in radio range units, which means that nodes in zone with a size less than one are under coverage of each other. Fig. 4(a) presents the average zone size as a function of the number of landmarks. Different neighbor densities are represented. We see that whatever the density, increasing the number of landmarks reduces the average zone size. This reduction is, however, more important in high dense networks. An important result here is that increasing the density when more than six landmarks are employed is useless if it is initially set to twenty.

Fig. 4(b) recasts the data in comparative forms. It shows the benefit, in percentage, on zone size for scenarios with $N$ landmarks with respect to the initial case in which only 3 landmarks are deployed. It thus expresses the cumulated reductions of zone size resulting from the addition of $(N-3)$ landmarks, represented as a relative value. We first observe that using ten landmarks, instead of three, reduces zone size by about 40 % in the sparse networks considered and by up to 65 % in the densely populated ones. Adding a fourth landmark is sufficient to reduce the zone size by approximatively 20 % when density is low, and 30 % when it is high.

Fig. 5 presents zone size as a function of density, for different numbers of landmarks. Clearly, we notice that the density effect depends on the number of landmarks used. With six or fewer landmarks, the average zone size grows with density, while this tendency appears to be reversed but with low significance with seven or more landmarks. For a sufficiently high starting density (at least twenty neighbors per node), an increasing density either does not reduce significantly zone size or provokes an adverse effect.

### 4.4.2 Maximum zone size

The maximum zone size is, as described above, the size of the largest zone in the whole network. It represents the largest distance, in number of hops, between any two nodes sharing the same coordinates. The distribution of maximum zone size is shown in Fig. 6. The first observation is that the density has a non-negligible effect on this measure, contrarily to the average zone size. While increasing the density reduces the maximum zone size, adding landmarks reduces it drastically. The most significant result is obtained with six and ten landmarks in relatively high dense networks. With six landmarks the maximum zone size is usually around one radio range, while it is less than one in almost all the cases when the number of landmarks is ten. This result is very important as it shows that nodes sharing the same coordinates are under the coverage range of each other.

Fig. 7(a) shows the same result. If we except networks with a density of ten neighbors per node, the curves are pretty close. This means that, beyond a certain threshold, increasing the density only has a limited effect on the average maximum zone size. Adding landmarks, however, significantly reduces this value. The gain obtained with one more landmark is considerable as shown in Fig. 7(b). The greatest gain is obtained when the first landmark is added. Indeed, using four landmarks instead of three reduces the maximum zone size by about 30 %. Continuing to add landmarks improves the result but in a less significant way.

From these results, we can first observe that using three landmarks in a network with a density of ten neighbors per node results in a 2.4 'radio range units'-wide largest zone. This means that there exist nodes sharing the same coordinates with, at least, three hops between them.

In such a case, using a 2-hop proactive ad hoc routing protocol coupled with a position-based routing protocol, as proposed by Caruso et al. [10], may be insufficient.

Now, by considering only networks with a neighbor density of twenty or more, we can see that we can avoid the use of a proactive routing protocol if we use seven landmarks or more, while the maximum zone size is less than or equal to one radio range. In this case, two nodes sharing the same coordinates are immediate neighbors. We recall that the goal of coupling a position based routing protocol to a proactive one is to allow communication between nodes within the same zone even if they are not directly connected.

### 4.4.3 Intra-zone distance

As described in Section 4.3, this measure represents the average distance between two nodes sharing the same virtual coordinates. The perfect coordinate system we can obtain is the one where every intra-zone distance is less than one radio range unit.

If we observe the results presented in Fig. 8, we notice that the intra-zone distance is usually less than one radio-range unit, especially when six landmarks or more are used. For this measure, as for the zone size one, the density has almost no effect.

We can see in Fig. 9(a) the average intra-zone distance as a function of the number of landmarks, for different neighbor densities. The tendencies are comparable with those of maximum zone size (cf. Section 4.4.2); except the case of ten neighbors per node, density does not seem to have a significant effect on the average intra-zone distance.

Fig. 9(b) presents the gain compared to the initial scenario and confirms the results of Section 4.4.2: adding one landmark reduces significantly the intra-zone distance while increasing the density has a less noticeable effect.

### 4.4.4 Number of nodes per zone

The number of nodes per zone is the average number of nodes sharing the same coordinates. While more nodes with the same coordinates means more ambiguity, it is obvious that the lower the number of nodes per zone, the more accurate the coordinate system.[2] We see in Fig. 10 that increasing the density, by adding nodes in the topology, increases the number of populated zones: zones with up to twenty nodes appear when density is set to fifty and only three landmarks are used.

In Fig. 11(a), we see that increasing the number of landmarks has a limited effect in sparse networks, in which there are, on average, fewer than three nodes sharing the same coordinates in all cases.

In more densely populated networks, however, the number of landmarks plays an important role, as it allows, as shown on Fig. 11(b), a reduction in the number of nodes per zone of up to 60 %. Also, when using ten landmarks, less than three nodes share one position, regardless of the density. In contrast, for a given landmark count, an increase in the neighbor density results in more populated zones, as shown in Fig. 11(c).

[2] Ambiguity is the unclearness due to having more than one node for one virtual position.

#### *4.4.5 Number of zones*

Fig. 12 presents the average number of zones in the network: first for any fixed density, as a function of the number of landmarks (Fig. 12(a)), and second for any fixed number of landmarks, as a function of neighbor density (Fig. 12(b)). The results on these figures are related to the results of Fig. 11. First, let us consider a fixed landmark count scenario, and analyze Figs. 11(c) and 12(b). We observe that, as neighbor density grows, both the number of zones and the number of nodes per zone grow. This implies that neighbor density growth adds ambiguity. Now, let us consider a fixed density scenario, and analyze Figs. 11(a) and 12(a). With the number of landmarks, the number of zones first tends to grow to a maximum value, but then decreases. Meanwhile, the number of nodes per zone steadily decreases. As a consequence, we identify the following behavior: up to the maximum value, the number of zones grows, but each zone contains fewer nodes. Beyond the maximum value, the number of zones regresses, and still each zone contains fewer and fewer nodes. In both cases, the distinction between nodes becomes more significant with the addition of landmarks.

## 5 Energy consumption concerns

As stated before, the goal of Jumps is to build a coordinate system based on hop-distances to landmarks. In order to allow each node to discover its coordinates, each landmark is supposed to flood the network with a probe, whose goal is to permit other nodes to discover their hop-distance to the landmark. Thus, every single landmark added requires one additional flood. In this section, we show, through a theoretical analysis, that in spite of increasing the number of floods, Jumps does not require more energy than other approaches that use only three landmarks to build an accurate coordinate system.

Let us consider a network composed of $\mathcal{M}$ nodes, where the average number of neighbors per node is $d_{neig}$ and the density per coverage range is $d_{couv}$ (cf. section 4.3). We suppose that there are $N$ landmarks and that we wish to discover the distances of each node to every landmark. For this purpose, each landmark floods the network with a probe.

In this scheme, each node should, at least, forward the probe once. If we assume a perfect communication scheduling and no collision nor packet loss, such a flooding results in $\mathcal{M}$ packets emissions and approximatively $\mathcal{M} \times d_{neig}$ packet receptions. For $N$ landmark, the distances discovery phase requires $\mathcal{M} \times N$ packet emissions and $\mathcal{M} \times N \times d_{neig}$ packet receptions.

In order to evaluate the energy consumption of this process, we consider the datasheet of the CC2420 radio chipset [20] present in the Moteiv Tmote Sky [21] sensors nodes and in the Xbow [22] motes. This chip consumes 19.7 mA for reception and proposes different emission powers ranging from $-25$ dBm to 0 dBm.

If we consider, as reference, the lowest power level (-25 dBm), increasing the output power (in Watt) by a factor $\lambda$ results in: (i) a $\sqrt[\alpha]{\lambda}$ gain in range ($\alpha$ denotes the path loss exponent); and therefore (ii) an increase in the coverage area by a factor $(\sqrt[\alpha]{\lambda})^2$. Table 1 summarizes the values for a free-space propagation model [23] and relates them to the CC2420 energy consumption figures.

Interpolating the current consumption $I_{Tx}$ as a function of density gives $I_{Tx} \simeq \frac{\sqrt{d_{neig}/d_0}+15.338}{1.8709}$. With a constant voltage $U$, we can state that the energy consumption can be expressed as a function of the number of landmarks ($N$) and of the average number of neighbors ($d_{neig}$) by :

$$E \simeq U \times \mathcal{M} \times N \times \left( \frac{\sqrt{d_{neig}/d_0} + 15.338}{1.8709} + 19.7 \times (d_{neig}/d_0) \right) \cdot$$

Fig. 13 represents the energy required by Jumps, compared to a reference case of a minimal transmission power and a single landmark. The transmission levels have been chosen in order to fit values used in simulations. Therefore, the maximum power level used leads to a network density of 51 nodes per coverage range. From these figures, we can see that increasing the transmission range can be equivalent to adding landmarks in terms of energy consumption. For instance, a scenario with 50 nodes as average number of neighbors and 3 landmarks is approximatively equivalent to a scenario with 10 neighbors and 10 landmarks.

Relating these results to the zone sizes' ones presented in section 4.4.1, a 50-node dense network with three landmarks leads to a zone size of 0.7 radio ranges and 7.7 nodes per zone, while a 10-node dense network with 10 landmarks leads to a 0.29 radio ranges zone size and 2.3 nodes per zone. Therefore, adding landmarks does not only improve accuracy of positioning, but also leads to either better energy consumption, allowing to reduce the transmission range of sensors, or to a lower network cost, allowing to deploy less sensors.

## 6 Conclusion and future works

In this paper, we presented JUMPS, a positioning system based solely on partial connectivity knowledge. Whereas existing algorithms use only three nodes as landmarks for hop-count positioning, we studied through extensive simulations the behavior of such a method with up to ten landmarks. The simulation results show that increasing the number of landmarks makes our coordinate system significantly more accurate than previous approaches. With a sufficient number of landmarks, the average distance between two nodes sharing the same coordinates is less than one radio coverage range, which makes it possible to avoid the use of any proactive routing protocol in favor of a pure position-based one. While adding one landmark involves one more flood when building the coordinate system, it avoids the need of increasing the neighborhood density, either by increasing the node population or the transmission power of these nodes. Thus, we show that adding landmarks, in addition to improving the accuracy of the positioning system, also reduces the energy required to establish an accurate coordinate system. Moreover, we believe that using more landmarks allows more freedom in landmark position choice, which gives ability to select them based on local criteria. In this paper, we consider only static sensor networks.

Table 1
This table gives some values of current power consumption ($I_{Tx}$), radio range coverage, and average neighbor density as a function of the output power. $R_0$ and $d_0$ represent the values for the minimum output power.

| Output power (dBm) | $I_{Tx}$ (mA) | Radio range $\alpha = 2$ | neighbors $\alpha = 2$ |
|---|---|---|---|
| -25 | 8.5 | $R_0$ | $d_0$ |
| -15 | 9.9 | $\sqrt{10} \times R_0 \simeq 3.16 \times R_0$ | $10 \times d_0$ |
| -10 | 11.2 | $10^{3/4} \times R_0 \simeq 5.62 \times R_0$ | $10^{3/2} \times d_0 \simeq 31.62 \times d_0$ |
| -7 | 12.5 | $10^{9/10} \times R_0 \simeq 7.94 \times R_0$ | $10^{9/5} \times d_0 \simeq 63.1 \times d_0$ |
| -5 | 13.9 | $10 \times R_0$ | $100 \times d_0$ |
| -3 | 15.2 | $10^{11/10} \times R_0 \simeq 12.59 \times R_0$ | $10^{11/5} \times d_0 \simeq 158.49 \times d_0$ |
| -1 | 16.5 | $10^{12/10} \times R_0 \simeq 15.84 \times R_0$ | $10^{12/5} \times d_0 \simeq 251.2 \times d_0$ |
| 0 | 17.4 | $10^{5/4} \times R_0 \simeq 17.78 \times R_0$ | $10^{5/2} \times d_0 \simeq 316.22 \times d_0$ |

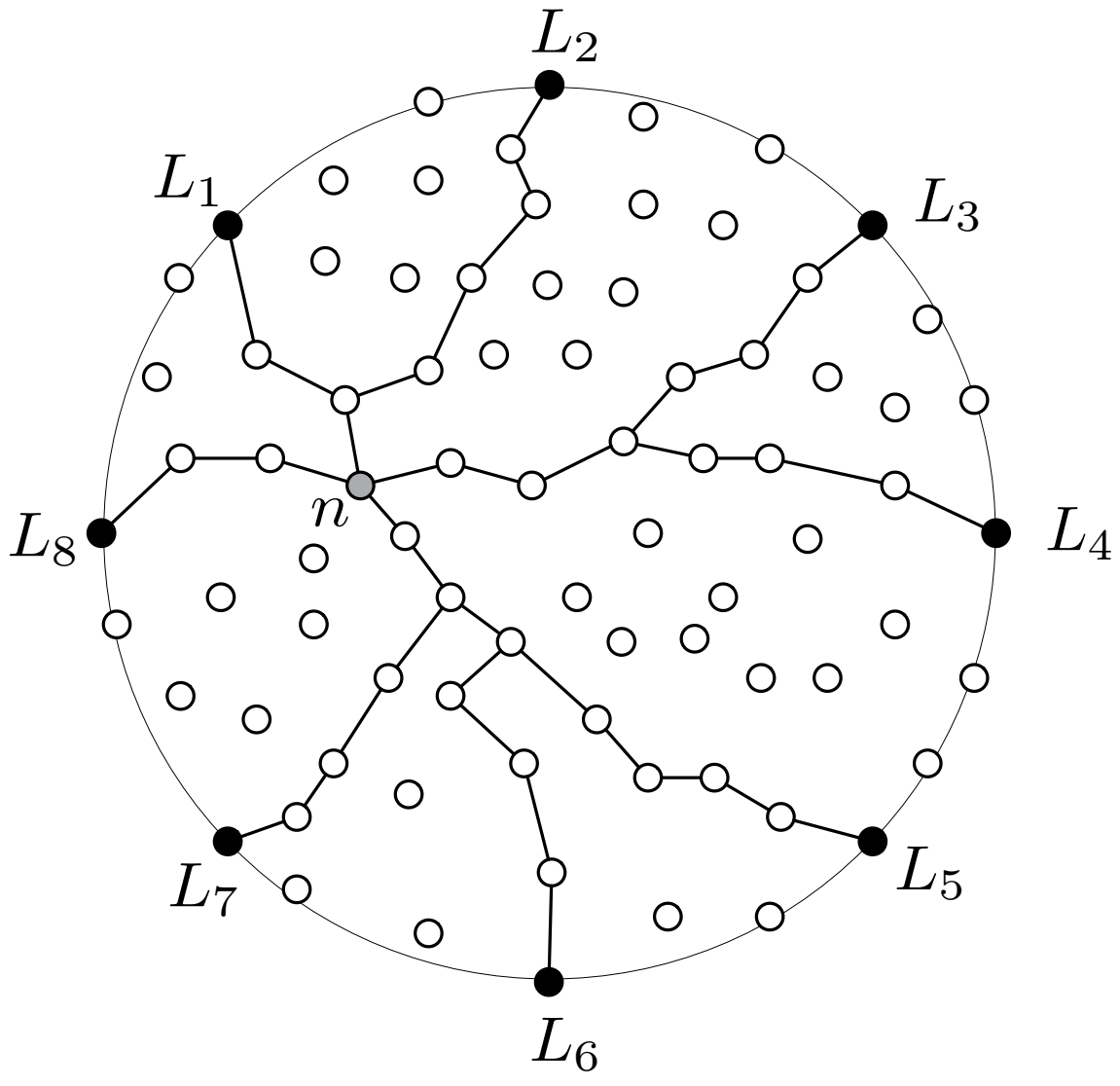

Fig. 1. Virtual coordinates computation with eight landmarks.

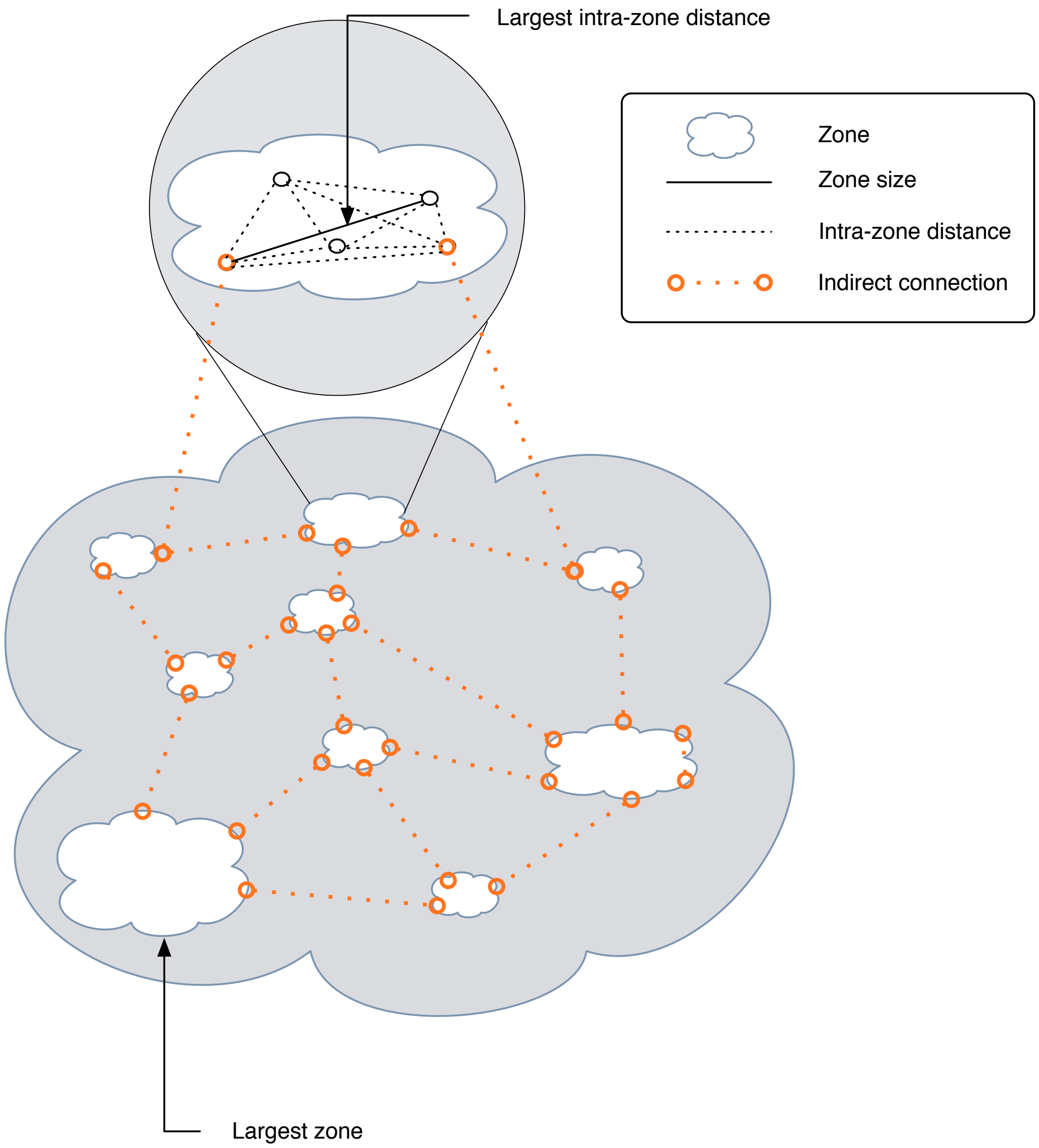

Fig. 2. Partition of the network into zones; details of a zone.

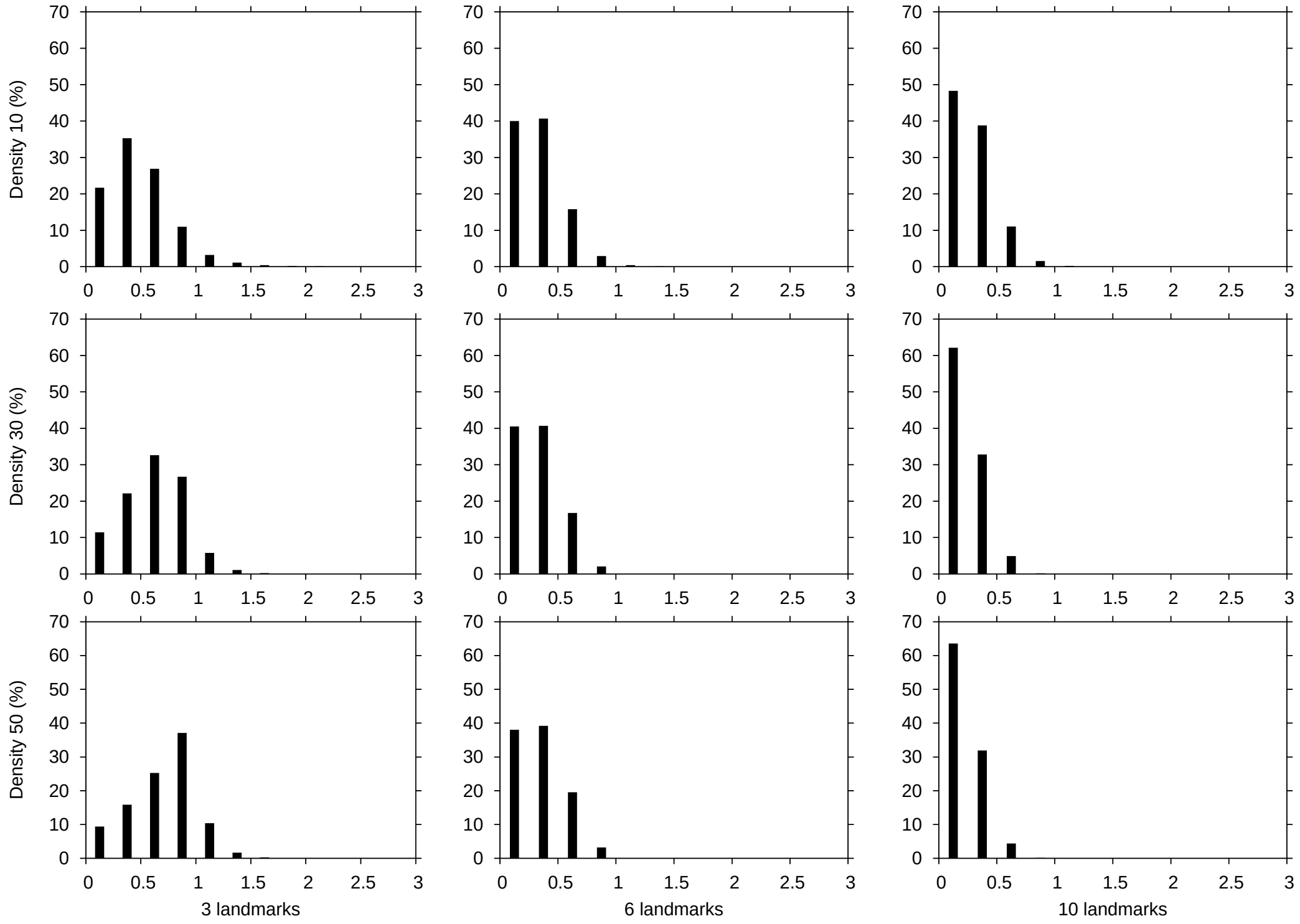

Fig. 3. Zone size distribution

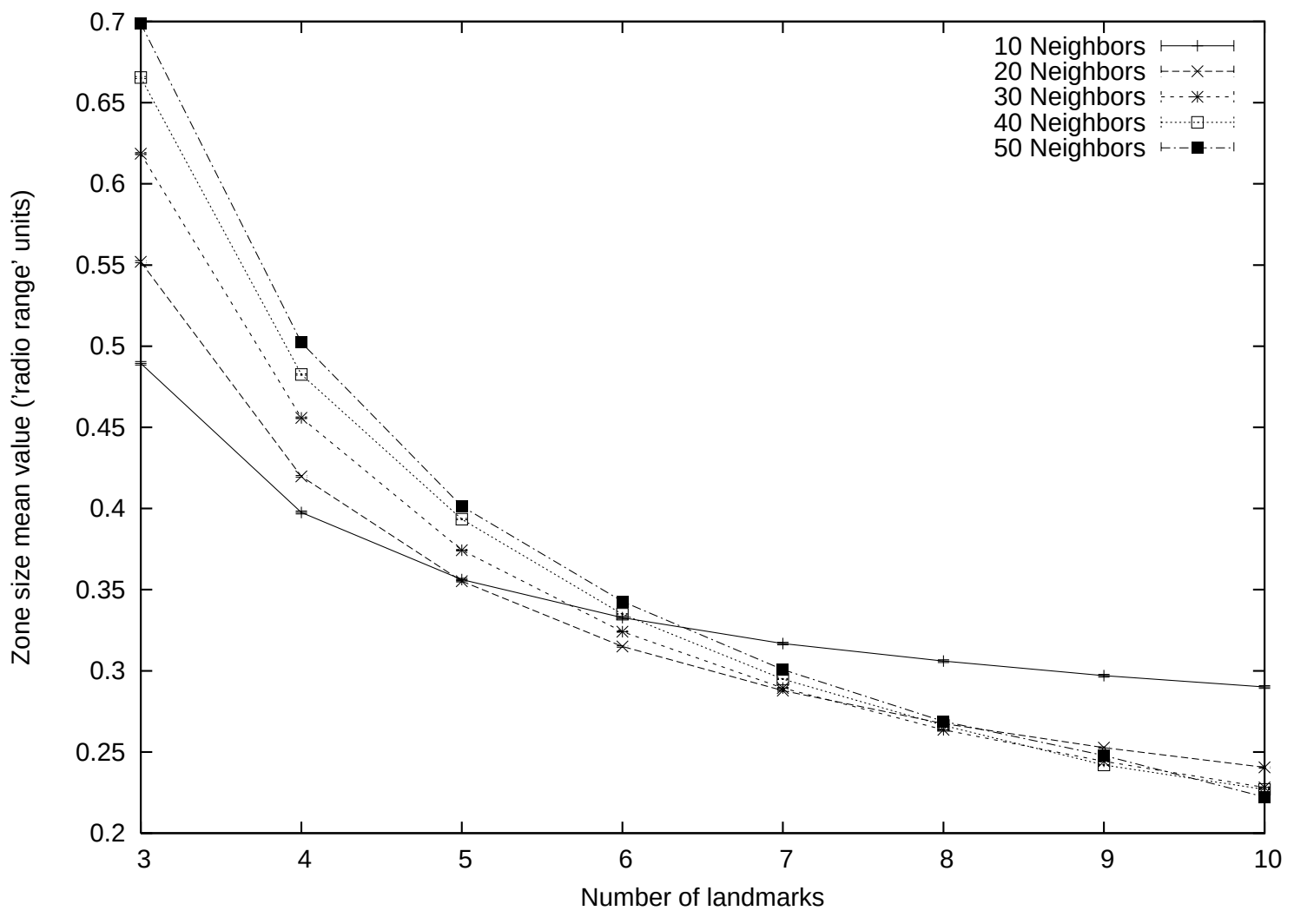

(a) Absolute value, in units of the radio range.

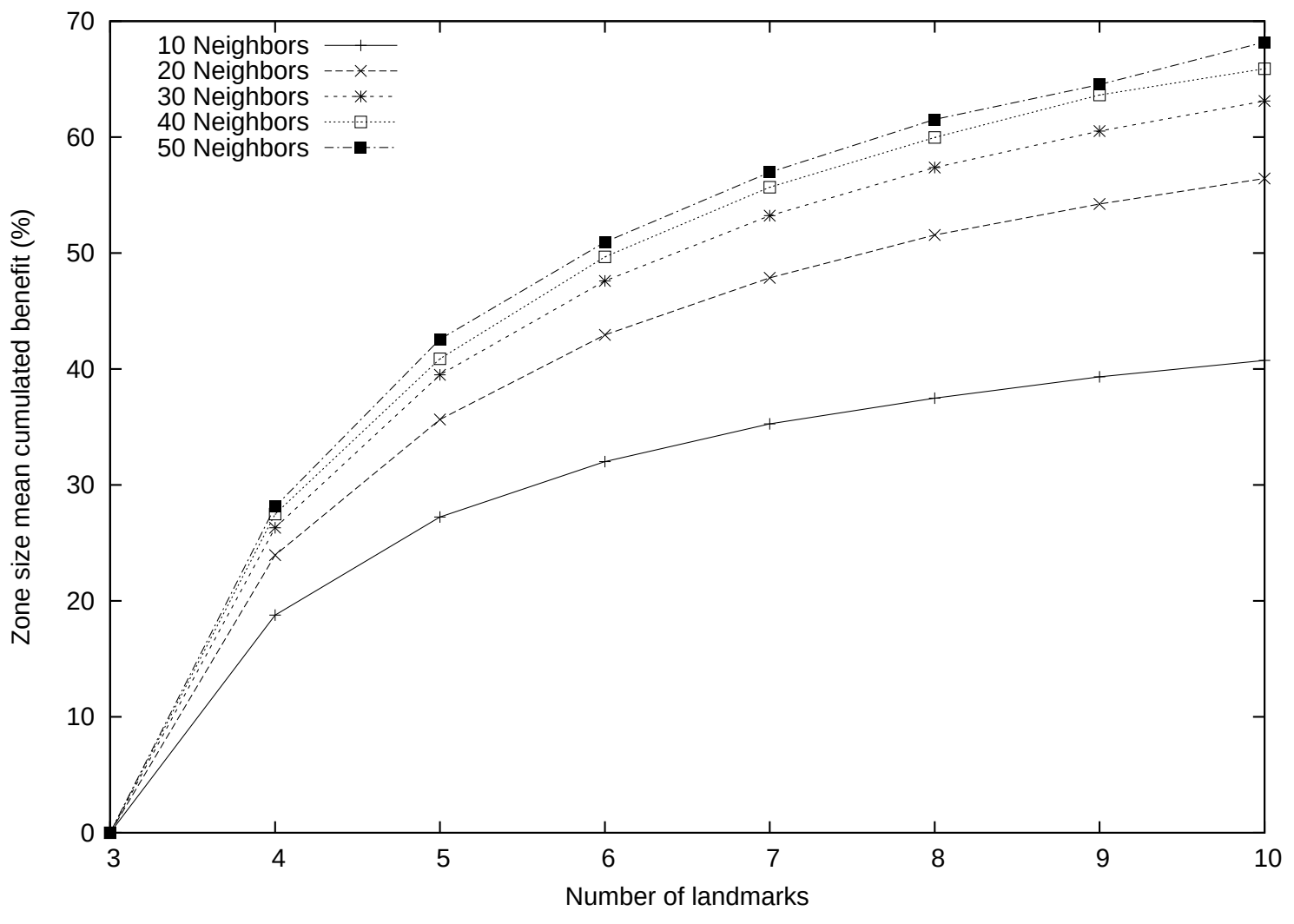

(b) Relative benefit (in percentage), with respect to the three landmarks scenarios.

Fig. 4. Zone size mean value in simulated scenarios, as functions of the number of landmarks, for five different neighbor densities.

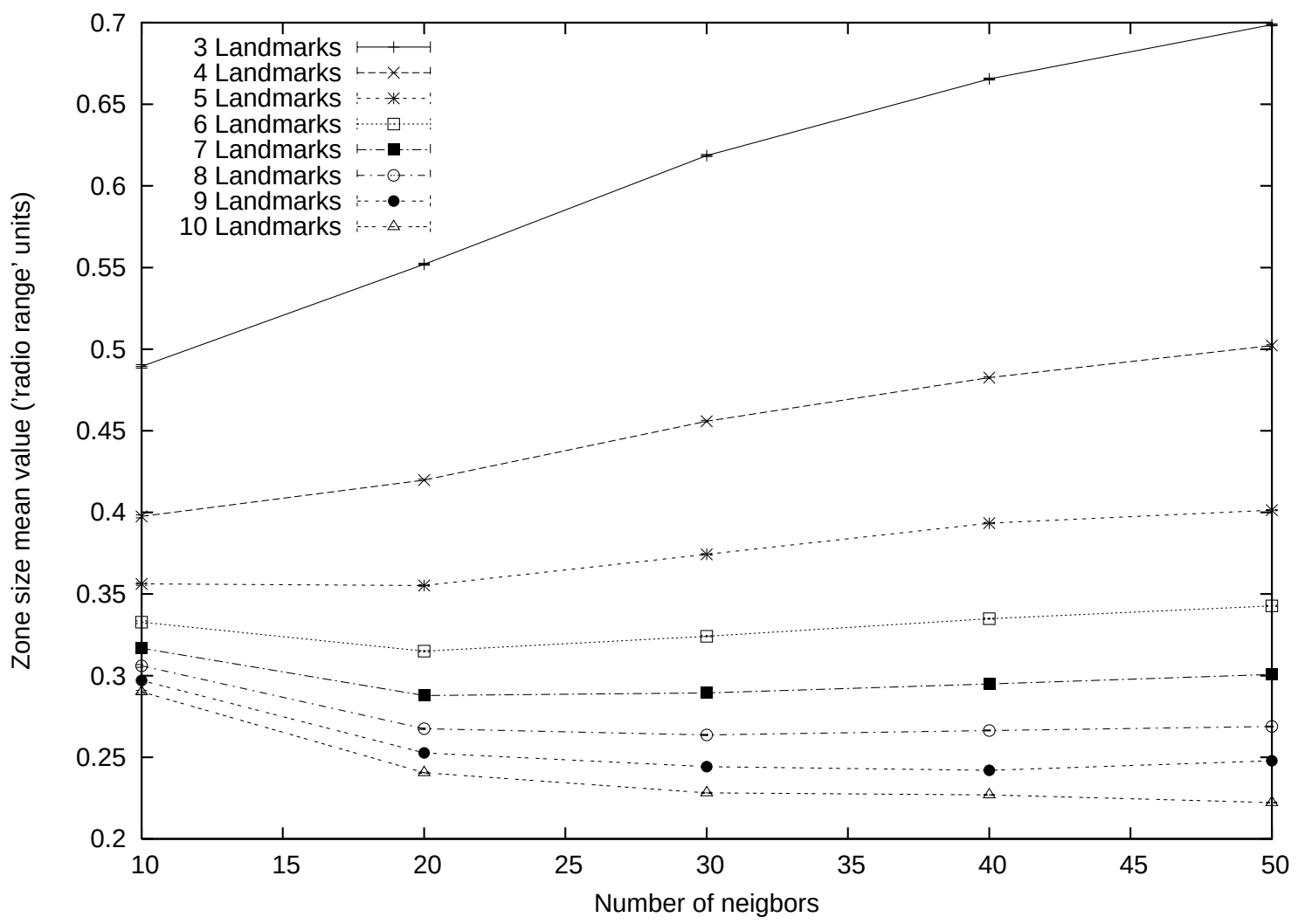

Fig. 5. Average zone size as a function of density. Notice how average zone size decreases when number of landmarks increases, whatever the density.

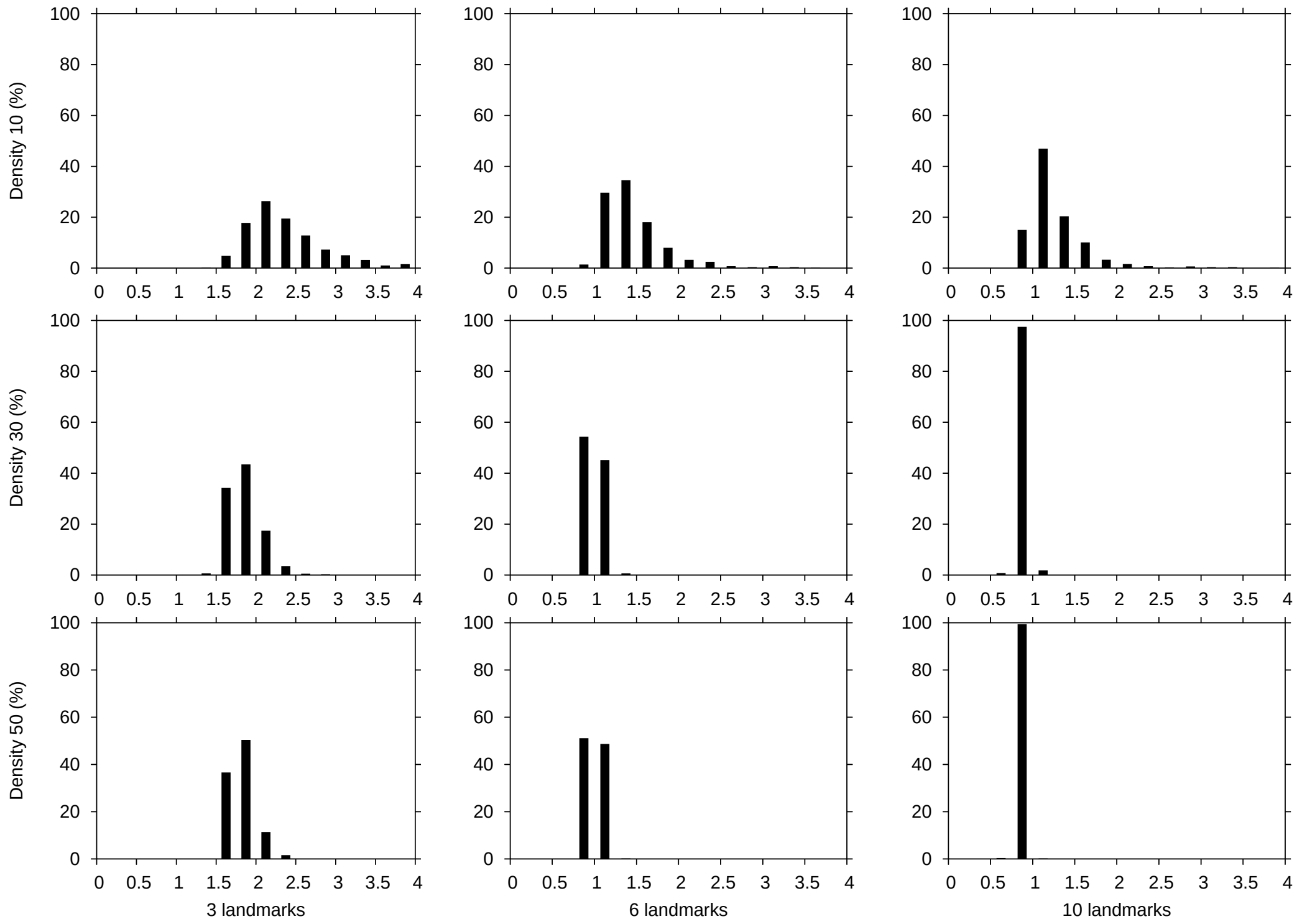

Fig. 6. Maximum zone size distribution.

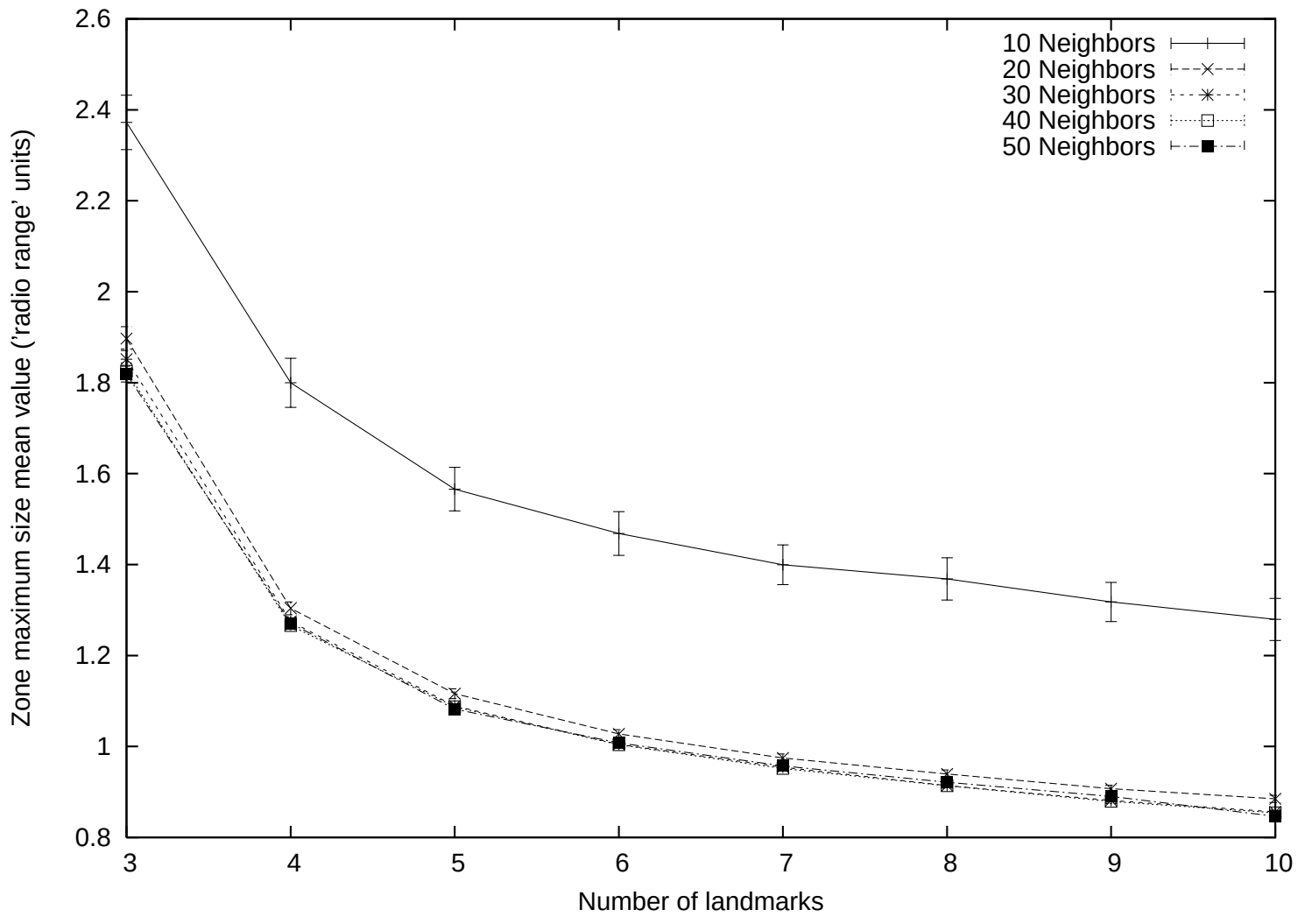

(a) Average maximum zone size.

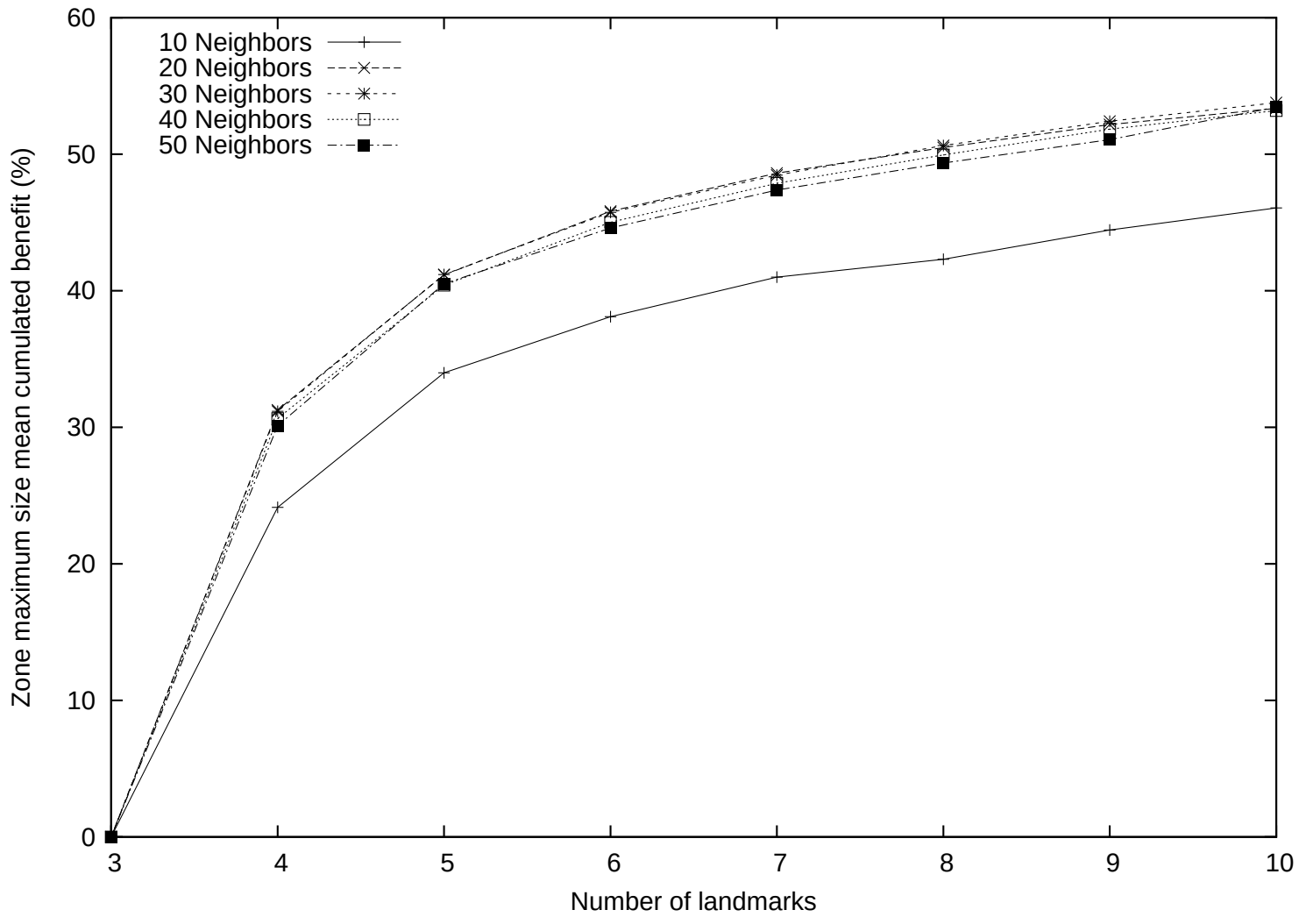

(b) Relative benefit (in percentage), with respect to the three landmarks scenarios.

Fig. 7. Maximum zone size as a function of number of landmarks. Observe that densities 20 and more have the same result.

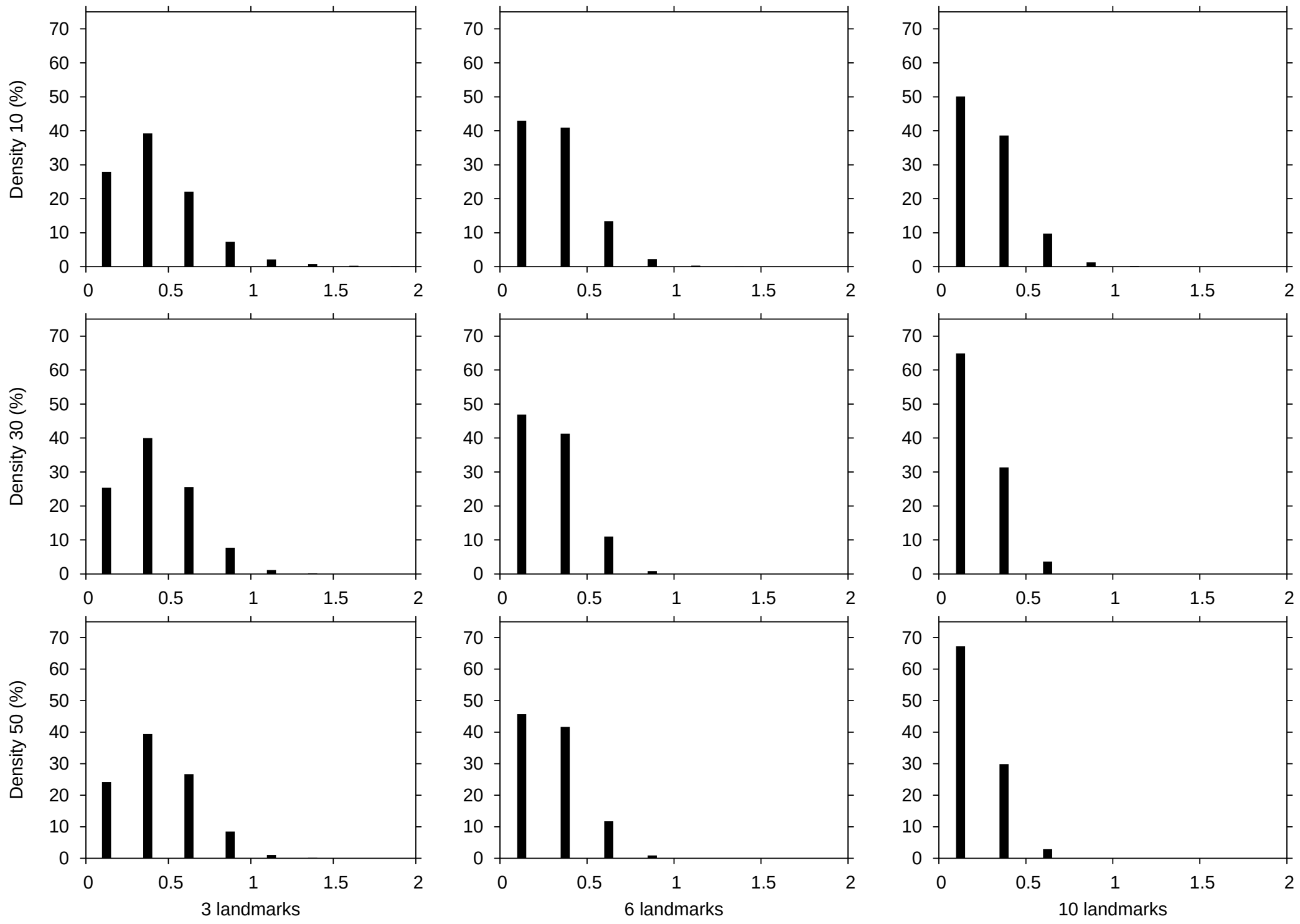

Fig. 8. Intra-zone size distribution.

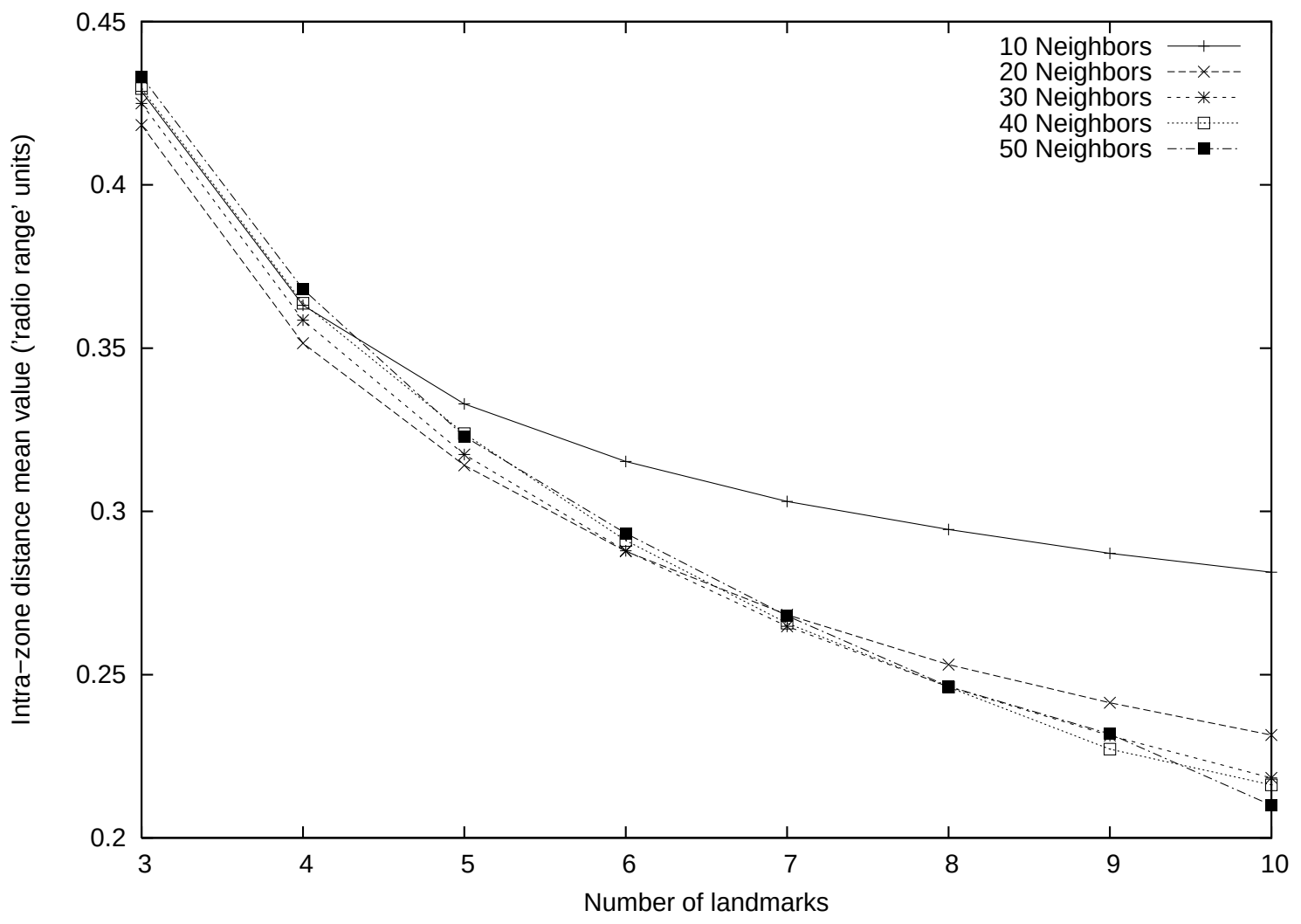

(a) Average intra-zone size.

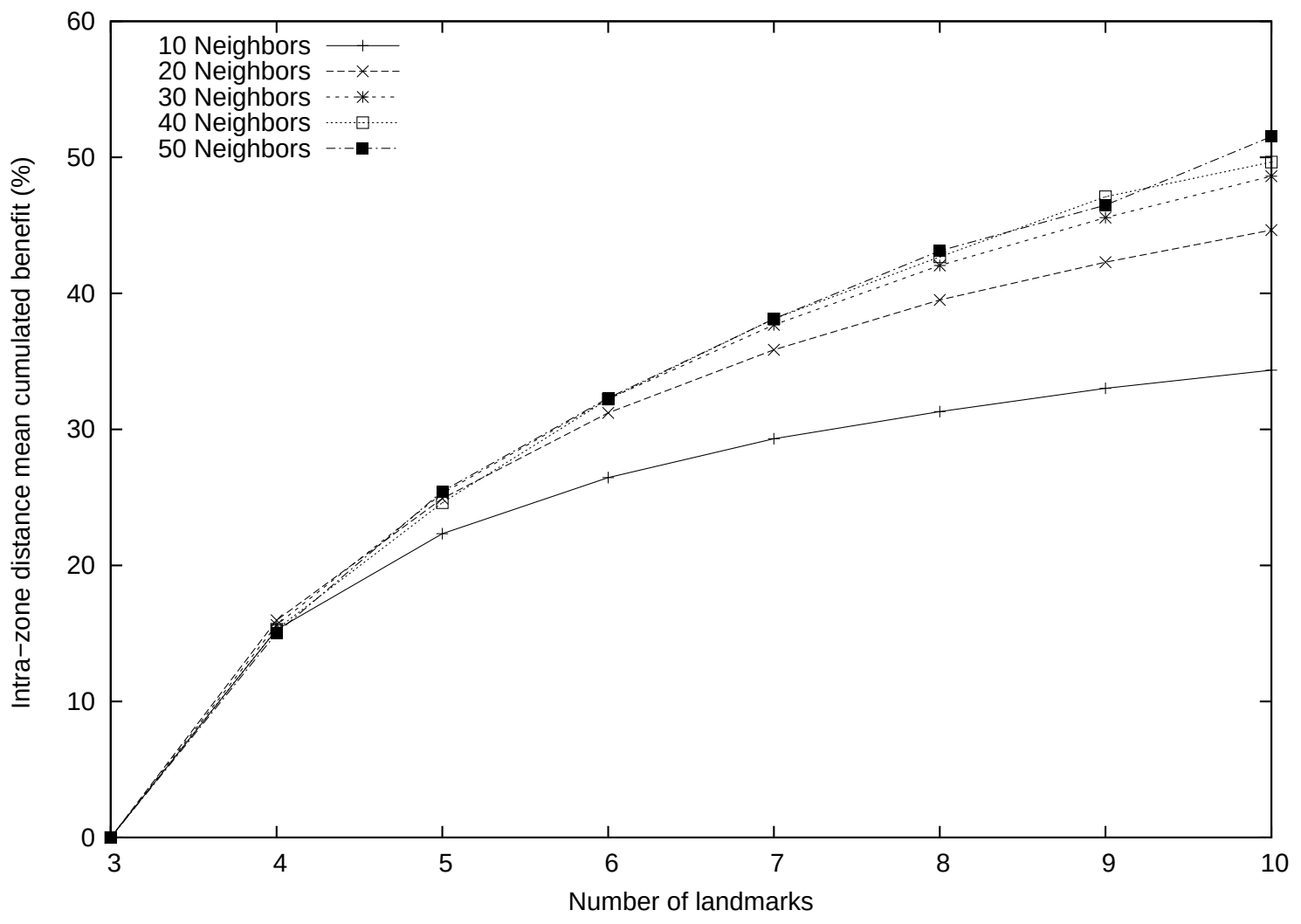

(b) Relative benefit (in percentage), with respect to the three landmarks scenarios.

Fig. 9. Intra-zone size as a function of number of landmarks. As for the maximum zone size, increasing the density beyond 20 neighbors per node has no effect.

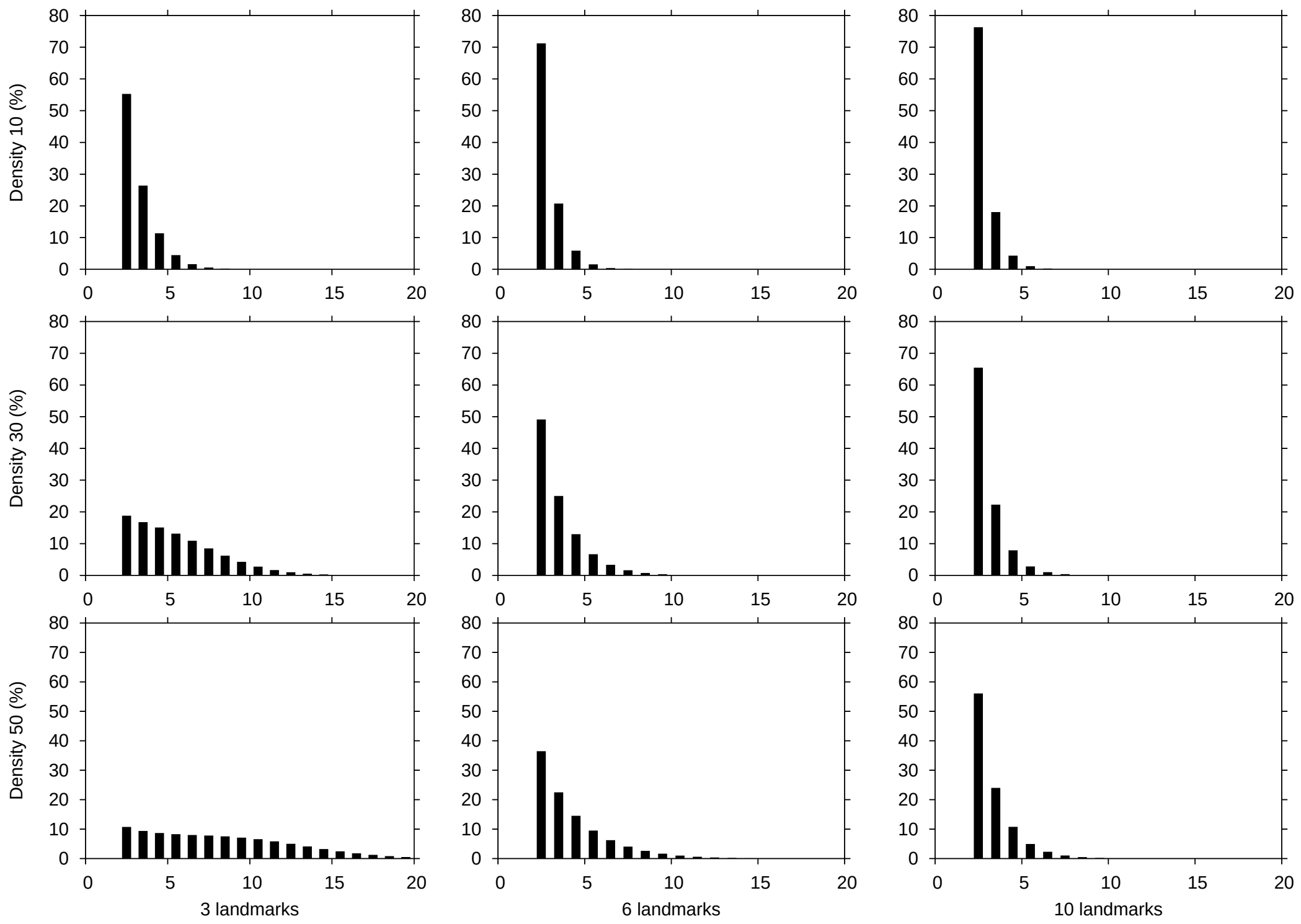

Fig. 10. Number of nodes per zone distribution.

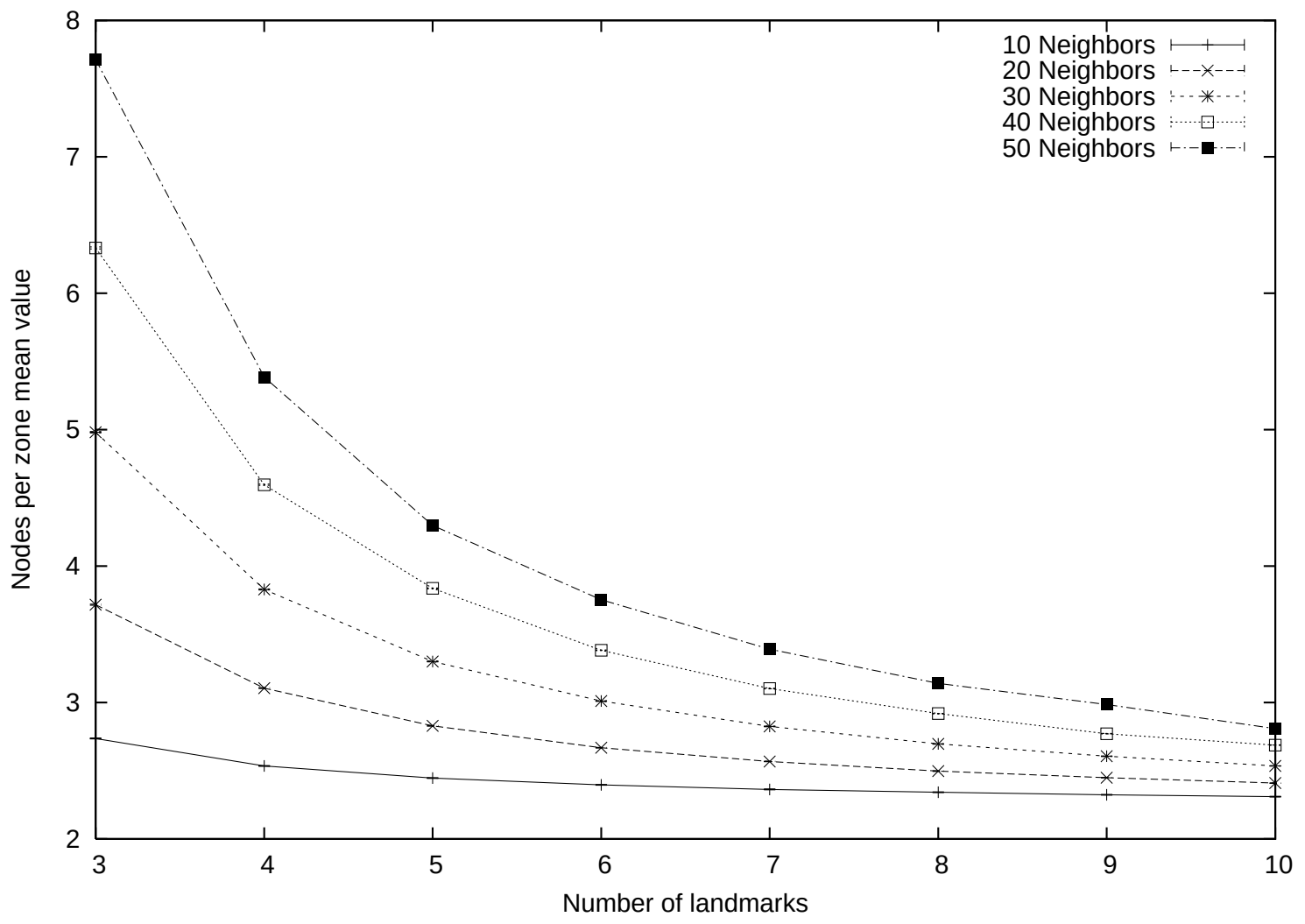

(a) Average number of nodes per zone.

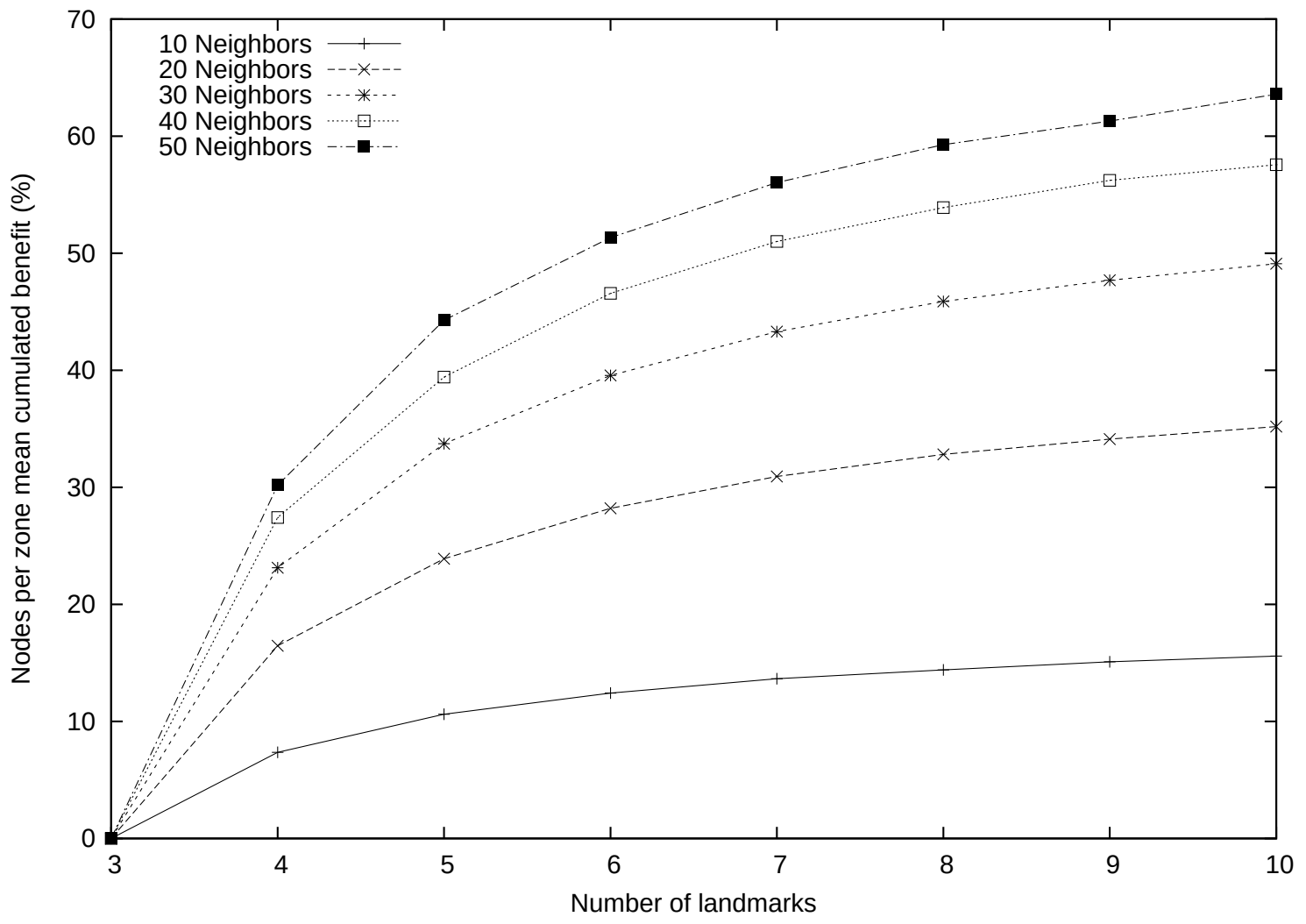

(b) Relative benefit (in percentage), with respect to the three landmarks scenarios.

Fig. 11. Number of nodes per zone as a function of number of landmarks.

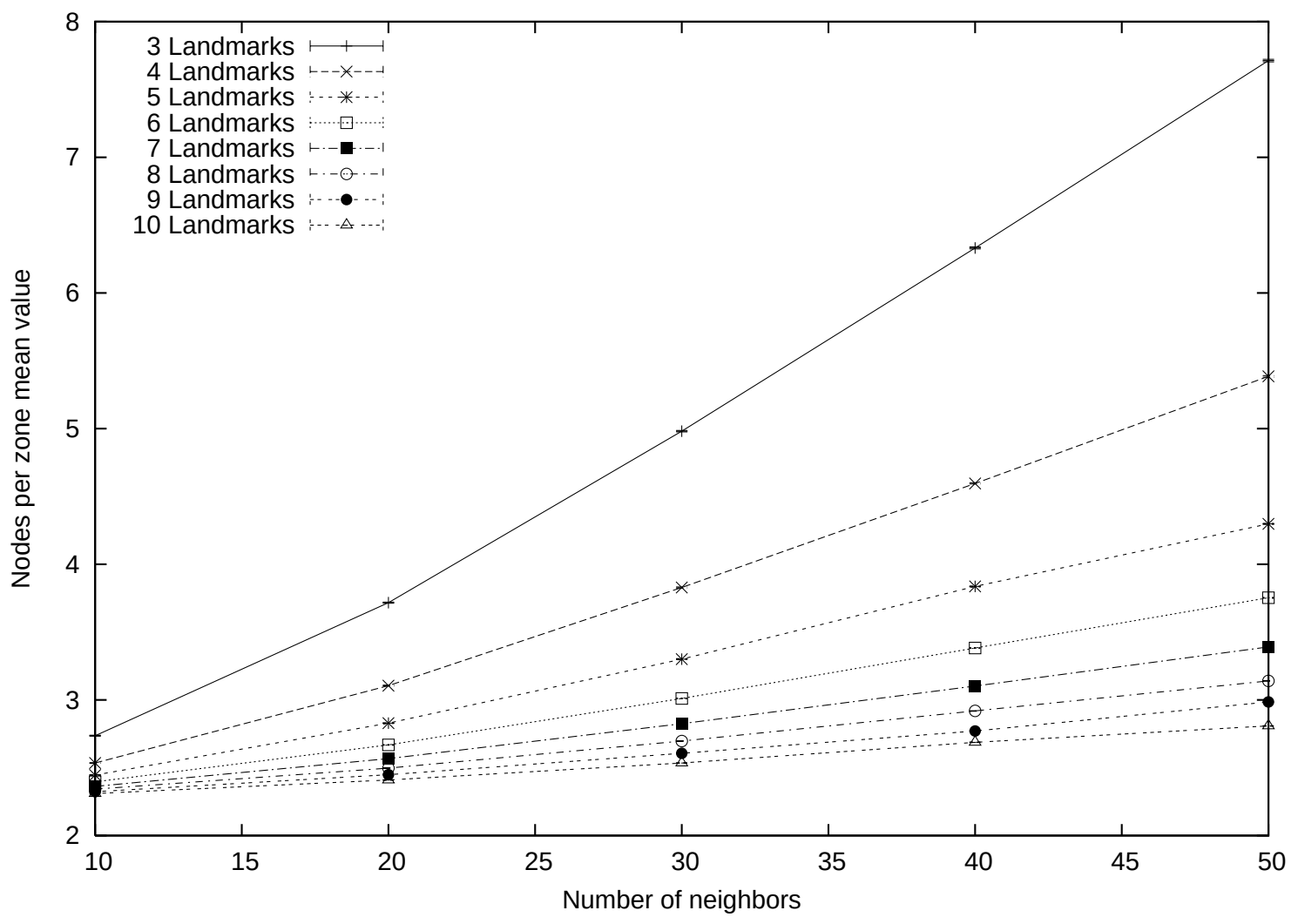

(c) Absolute value, in units.

Fig. 11. Nodes per zone mean value in simulated scenarios, as a function of neighbor density, for height different numbers of landmarks.

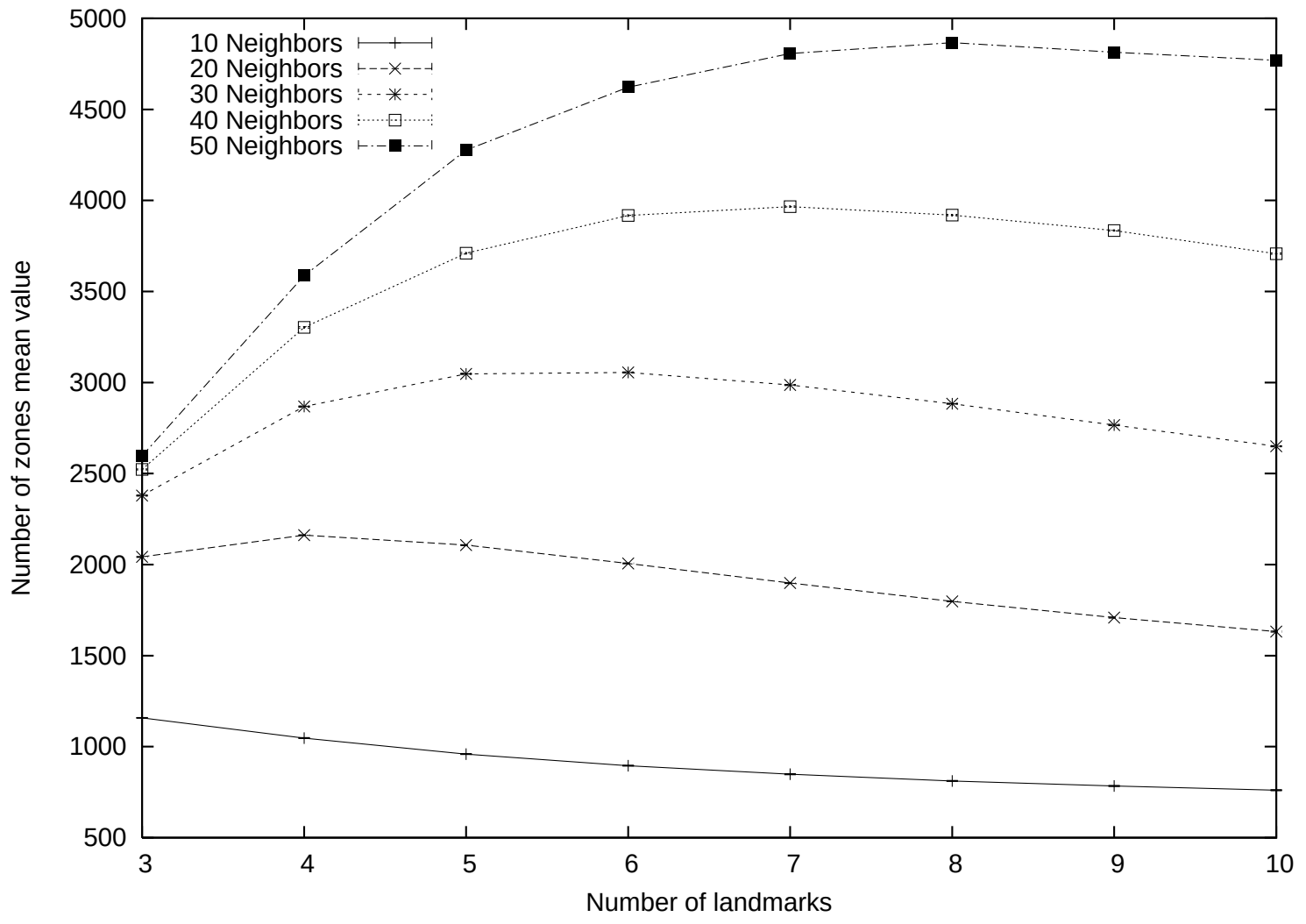

(a) Absolute value, as functions of the number of landmarks, for five different neighbor densities.

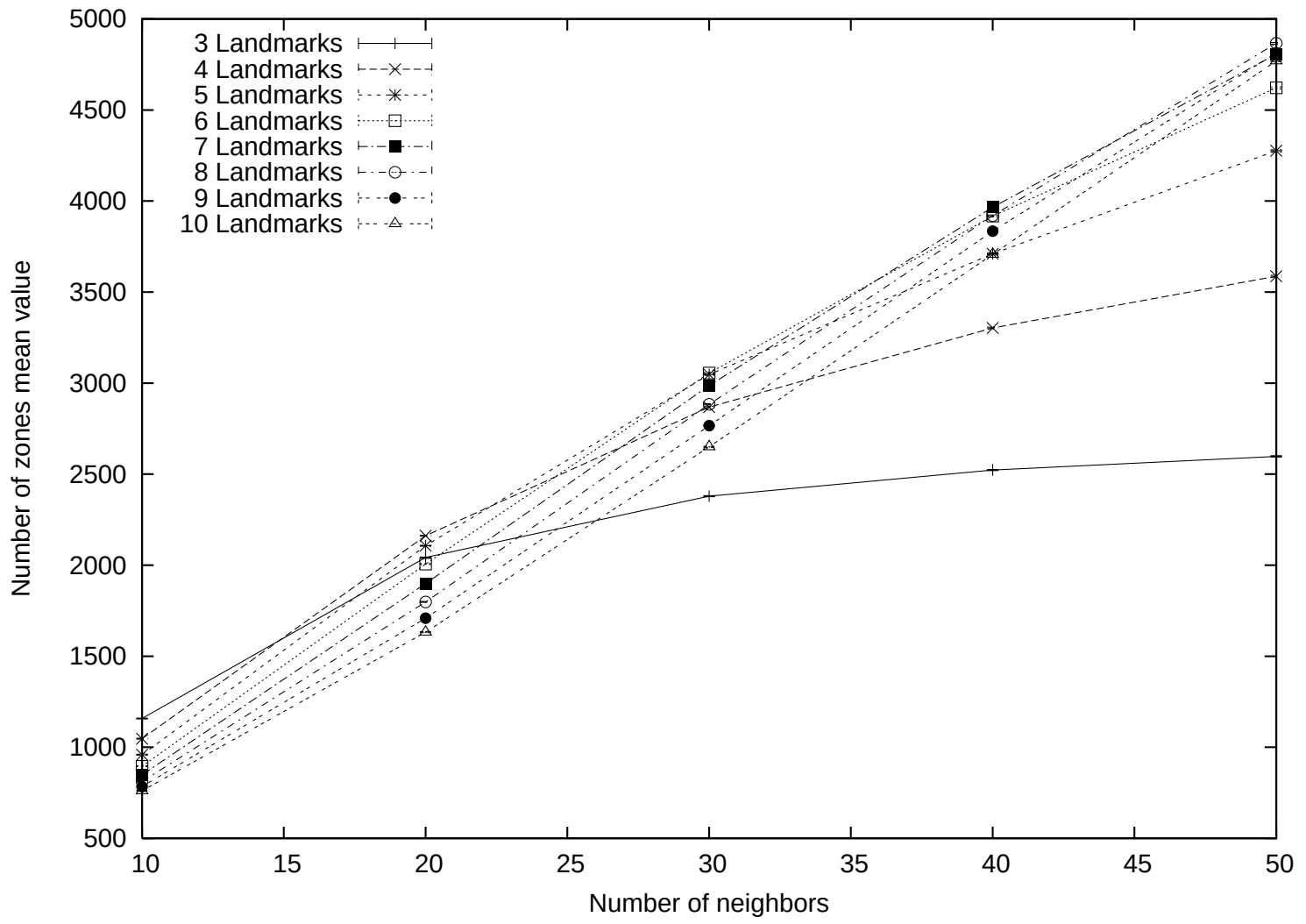

(b) Absolute value, as functions of the neighbor density, for height different numbers of landmarks.

Fig. 12. Number of zones mean value in simulated scenarios.

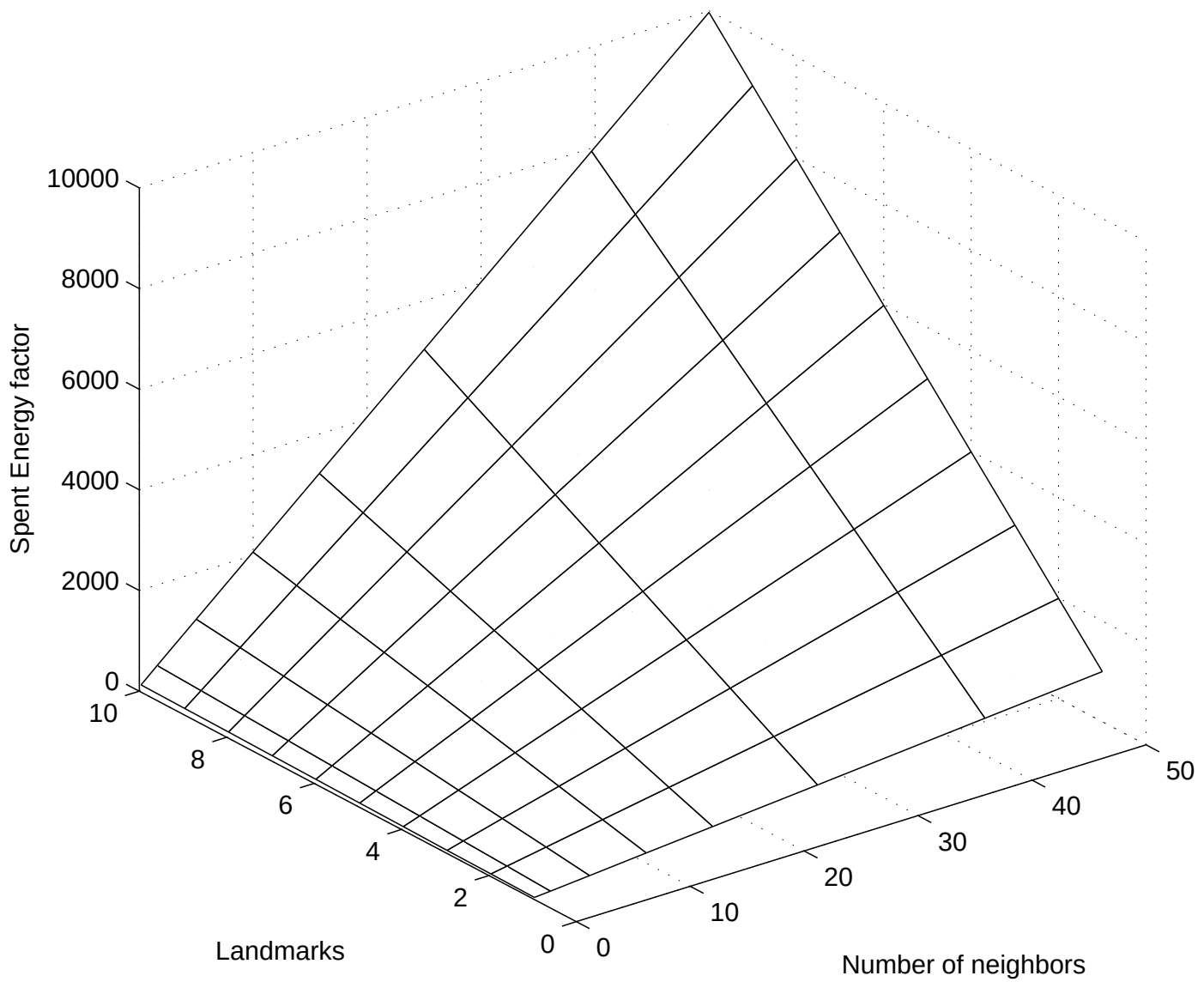

(a) We observe here that increasing the node density requires more energy that increasing the number of landmarks

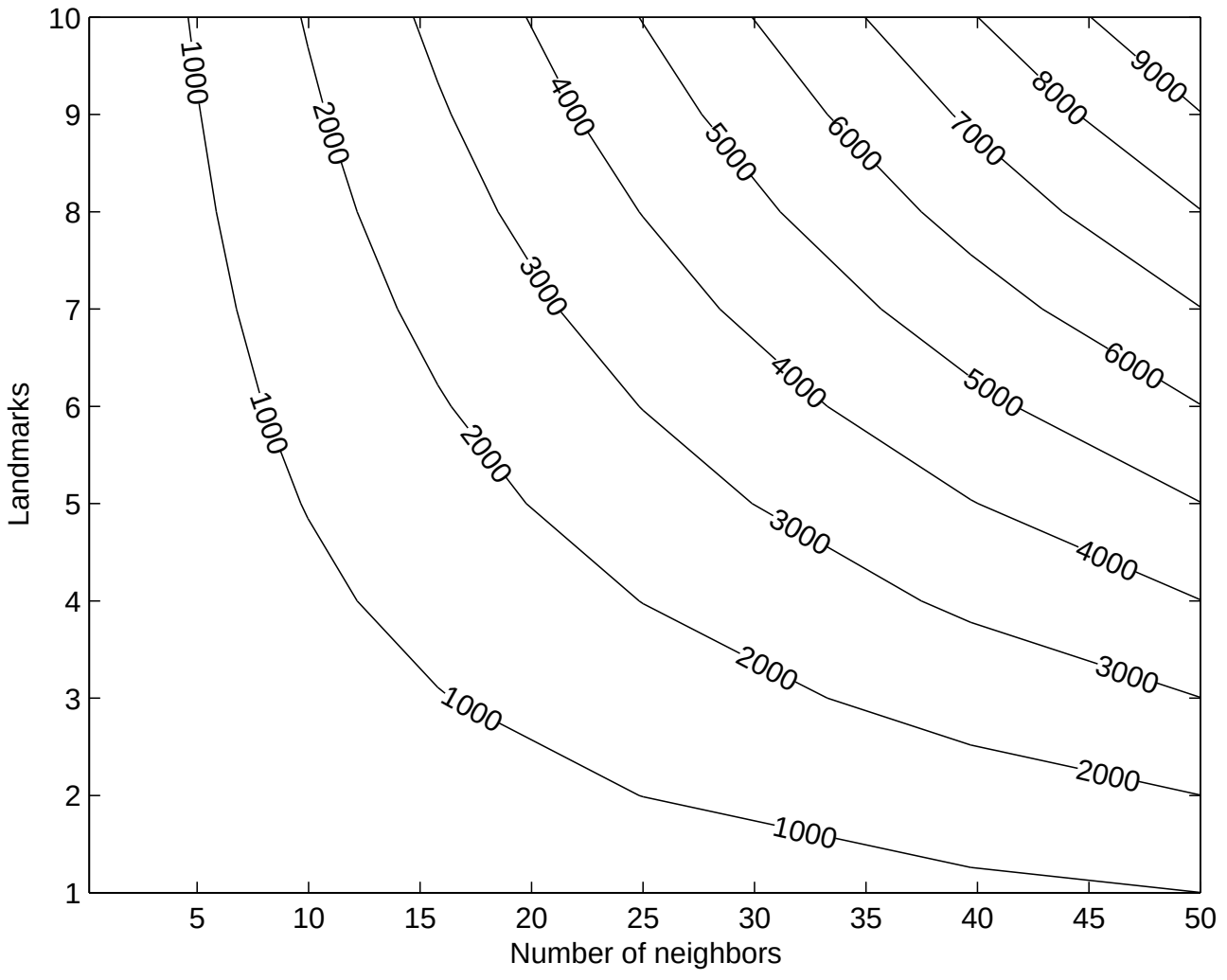

(b) Each curve in this figure represents the energy consumption as a factor of the reference level (one landmark with minimum transmission power).

Fig. 13. Energy consumption per node as a function of number of landmarks and neighbor density. The results are represented as a comparison to an initial case of a single landmark with minimum transmission power.